\newlength{\intwidth}

\documentclass{jfm}

\usepackage{amssymb}

\usepackage{array}
\usepackage{amsmath}
\usepackage[dvipsnames]{xcolor}
\usepackage{latexsym}
\usepackage{bezier}
\usepackage{psfrag,graphicx}
\usepackage{bm}
\usepackage{float}
\usepackage{multirow}
\usepackage{mathrsfs}
\usepackage{color}
\usepackage{enumerate}
\usepackage{natbib}
\usepackage{nicefrac}
\usepackage[colorlinks, citecolor=blue]{hyperref}
\usepackage{overpic}
\usepackage{rotating}
\usepackage{verbatim}

\begin{document}

\title[Spatial marching in curved channel flow]{Spatial marching with subgrid-scale local exact coherent structures in non-uniformly curved channel flow}

\author{
  Runjie Song\aff{1}
  \and Kengo Deguchi\aff{1}
}
 
 \affiliation{
   \aff{1}School of Mathematics, Monash University, VIC 3800, Australia
}

\maketitle

\begin{abstract}
We propose a novel multiple-scale spatial marching method for flows with slow streamwise variation. The key idea is to couple the boundary region equations, which govern large-scale flow evolution, with local exact coherent structures that capture small-scale dynamics. This framework is consistent with high-Reynolds-number asymptotic theory and offers a promising approach to construct time periodic finite amplitude solutions in a broad class of spatially developing shear flows. As a first application, we consider a non-uniformly curved channel flow, assuming that a finite-amplitude travelling wave solution of plane Poiseuille flow is sustained at the inlet. 
The method allows for the estimation of momentum transport and highlights the impact of the inlet condition on both the transport properties and the overall flow structure.
We then consider a case with gradually decreasing curvature, starting with Dean vortices at the inlet. 
In this setting, small external oscillatory disturbances can give rise to subcritical self-sustained states that persist even after the curvature vanishes. 
\end{abstract}

\section{Introduction}\label{sec:introduction}



Flows with slow spatial development, such as boundary layers, jets and wakes, are common in physical and engineering applications. Fully resolved direct numerical simulations of many practical problems are thus prohibitively expensive, necessitating the use of models for under-resolved dynamics. Yet, accurately quantifying the impact of subgrid-scale processes on large-scale flow remains a fundamental challenge in fluid dynamics, largely due to the limited understanding of coherent structures. 

The investigation of coherent structures in near-wall turbulence has progressed through the computation of minimal flow units in simple parallel shear flows \citep[e.g.,][]{Jimenez_Moin_1991,hamilton1995}. To date, the most rational theoretical description of these structures is based on exact coherent structures (ECS) and their asymptotic analysis in the high-Reynolds-number limit. ECS are ubiquitously found in parallel shear flows \citep{Nagata_1990,Clever_Busse_1992,faisst2003,wedin2004}; they are non-chaotic statistically steady states such as travelling waves or periodic orbits. 
Physically, ECS in shear flows represent the simplest states that capture the self-sustaining process of coherent structures via the interaction of rolls, streaks, and waves \citep{waleffe1997,wang2007}. This process was later shown to be consistent with the vortex–wave interaction theory of \cite{hall1991}, a large Reynolds number asymptotic framework derived from the Navier–Stokes equations \citep[see also][]{HALL_SHERWIN_2010, deguchi2014,Brand_Gibson_2014} and is therefore referred to as SSP-VWI throughout this paper.

Although ECS are typically unstable and therefore difficult to observe directly in shear flows, they play crucial roles in understanding fluid motion from the perspective of dynamical systems theory. One key significance of ECS is that they provide a systematic framework for studying subcritical transition. Certain ECS can be observed in simulations or experiments when the flow is controlled to prevent transition to either turbulence or laminar flow \citep{Itano_Toh_2001,wang2007,Eckhardt2007,de_2012}.
Furthermore, at low Reynolds numbers, ECS frequently appear within chaotic dynamics \citep{hof2004,Kawahara_2012,Suri_2020}. When a sufficiently large number of periodic solutions are collected, they can be used to reconstruct the probability density function of the chaotic attractor \citep{W25}. Thus, ECS are often referred to as the skeleton of turbulence.

Over the past 30 years, the use of ECS has significantly advanced our understanding of shear flows within periodic domains.
However, extending the ECS framework to the aforementioned spatially developing flows remains a major challenge. In this study, we examine a curved channel flow with non-uniform wall curvature, which serves as a prototype for addressing this difficulty. This configuration corresponds to the infinite aspect ratio limit of the curved duct problem \citep{Haines_Denier_Bassom_2013}. A historical overview of this problem, originating from \citet{Dean_1928_channel}, is provided in \citet{Rigo_2021}. 

Previous studies have largely focused on cases with constant curvature. When the curvature is zero, the problem reduces to plane Poiseuille flow, where subcritical transition can be explained using ECS \citep[e.g.][]{{Itano_Toh_2001}}. In contrast, sufficiently strong curvature leads to a supercritical transition driven by centrifugal instability \citep{Dean_1928_channel}. 
When curvature varies along the streamwise direction, interaction between local distinct transition pathways within a single flow configuration may become possible. Such cases are frequently encountered in practical applications but have been the subject of very limited theoretical investigation. Dynamical systems theory would also offer a promising framework for analysing such complex flow configurations; however, a major challenge lies in the difficulty of computing ECS in large computational domains.

Capturing the spatial evolution of coherent structures in large computational domains has long been a challenging task, prompting the development of various strategies to reduce computational cost. 
Since the 1980s, it has been recognised that in slowly spatially developing flows, numerical integration in the streamwise direction is feasible by parabolising the governing equations. Pioneering numerical examples of this approach include the linear stability analysis of G\"ortler vortices by \cite{Hall_1983} and works summarised in \cite{Rubin_Tannehill_1992}. Among the various proposed methods, the boundary region equations \citep[BRE, see][]{Kemp_1951} have become one of the mainstream approaches in academic research \citep{Hall_1988, RICCO_WU_2007, Deguchi_Hall_Walton_2013, Xu_Zhang_Wu_2017,Rigo_2021}. However, the BRE approach has certain limitations. For instance, the inlet condition is typically restricted to a laminar base flow with small, slowly time-varying perturbations. In addition, the BRE formulation is inherently incapable of capturing the SSP-VWI, as the wave components responsible for sustaining turbulence are associated with the discarded fast-scale (or `subgrid-scale') dynamics.

This study combines the BRE and ECS methods to incorporate the effects of subgrid-scale SSP-VWI into spatial marching. We mainly consider the case in which the channel curvature gradually increases from zero. At the upstream end, we assume that the edge state (i.e. spatially local ECS) of plane Poiseuille flow is sustained.
The inlet condition therefore involves large-amplitude perturbations with fast-scale oscillations, which, to the best of our knowledge, have not been incorporated in previous spatial marching computations.

In the next section, we present the mathematical and computational formulation of the reduced problem. The computational setup and method are summarised in section~\ref{sec:3}, followed by numerical results in section~\ref{sec:4}. Our method yields time-periodic spatially global ECS of the reduced problem, which approximate those of the full Navier–Stokes equations. Finally, concluding remarks are given briefly in section~\ref{sec:5}.






\section{Formulation of the problem}\label{sec:2}



\subsection{Curved channel problem}\label{sec:2.1}

\begin{figure}
\centering
\begin{overpic}[width=0.95 \textwidth]{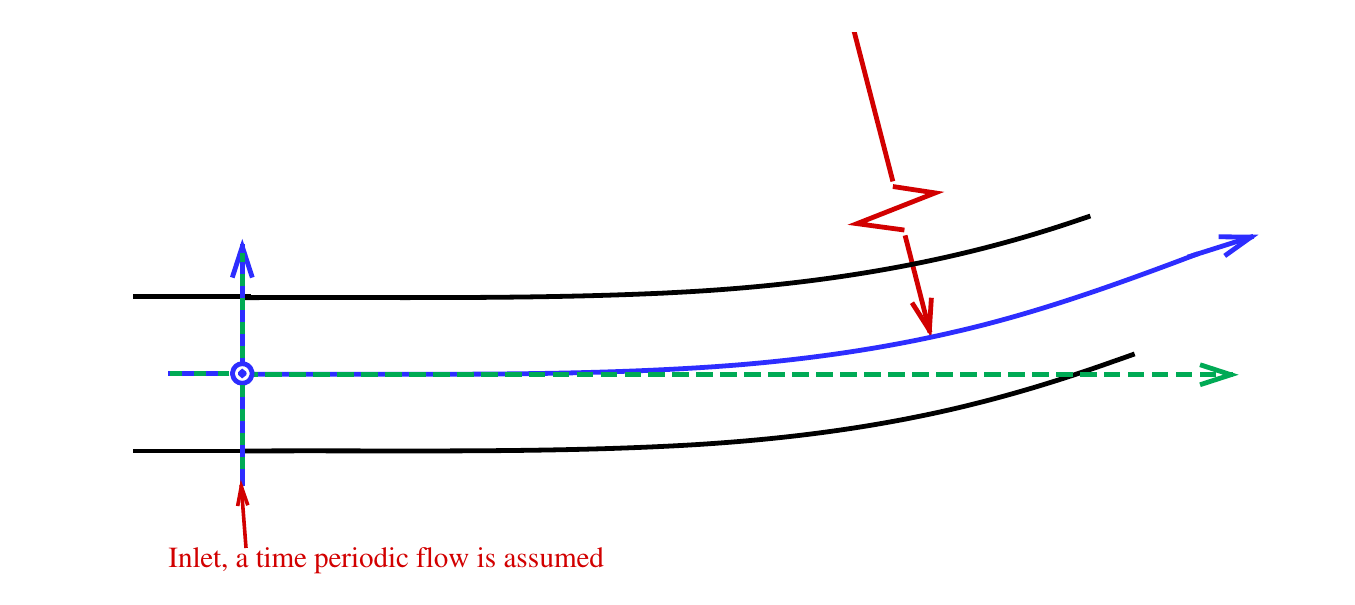}
\put(11,24){\textcolor{red}{$ y=1$}}
\put(9.5,13){\textcolor{red}{$ y=-1$}}
\put(15,30){\textcolor{blue}{$ y$}, \textcolor{ForestGreen}{$\check{y} $}}
\put(19,18){\textcolor{blue}{$ z$}, \textcolor{ForestGreen}{$\check{z} $}}
\put(92,17){\textcolor{ForestGreen}{$\check{x} $}}
\put(92,28){\textcolor{blue}{${x} $}}
\put(68,32){\textcolor{red}{$ a $}}
\end{overpic}
\caption{
Schematic of the curved channel flow problem. Length is nondimensionalised using the half-width of the channel. The body-fitted coordinates $(x,y,z)$ are attached to the channel centreline. For a given radius of curvature $a$, the centreline has a unique parametrisation in Cartesian coordinates $(\check{x},\check{y},\check{z})$.
}
\label{fig:sketch}
\end{figure}
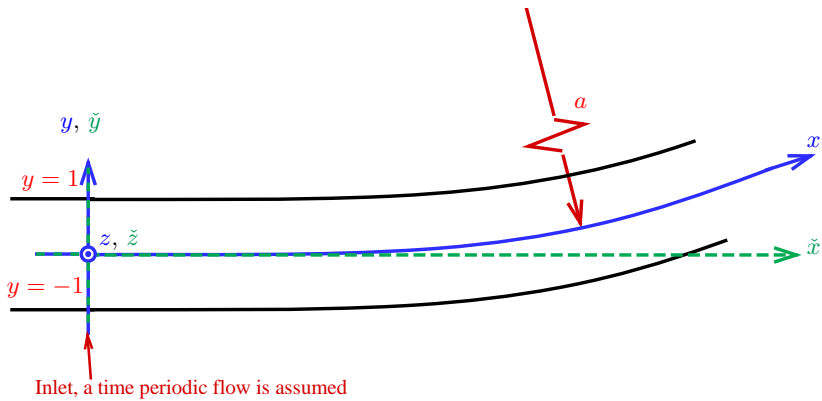

Consider an incompressible fluid with kinematic viscosity $\nu^*$ flowing through a non-uniformly curved channel of dimensional width $2d^*$. The channel curvature is specified by $a^*$, the dimensional local radius of curvature of the channel centreline. It is assumed that the channel is not curved in the spanwise direction. 
As shown in figure~\ref{fig:sketch}, we employ an orthogonal curvilinear coordinate system $(x^*, y^*, z^*)$ aligned with the geometry of the channel. Here, $y^*$ and $z^*$ measure distances along two orthogonal directions within the cross-section, while $x^*$ is defined as the arc-length coordinate along the channel centreline, starting from the inlet.
In other words, we adopt body-fitted coordinates  \citep{slattery1999} attached to the channel centreline.
To avoid confusion, Cartesian coordinates are denoted as $(\check{x}^*,\check{y}^*,\check{z}^*)$.


Using the half-width $d^*$ as the length scale, the bulk velocity $U_b^*$ as the velocity scale, and the fluid density as the density scale, the governing equations in the non-dimensional coordinates $(x,y,z)=(x^*,y^*,z^*)/d^*$ are written as
\begin{eqnarray}\label{NSeq}
(\partial_t +\mathbf{u}\cdot \nabla)\mathbf{u}=-\nabla p+R^{-1}\nabla^2 \mathbf{u},\qquad \nabla \cdot \mathbf{u}=0,
\end{eqnarray}
where $R=U_b^*d^*/\nu^*$ is the bulk Reynolds number.
The pressure gradient $\nabla p$ includes a component that ensures the non-dimensional bulk velocity remains unity.  
The velocity field can be expressed as $\mathbf{u}=u\mathbf{e}_x+v\mathbf{e}_y+w\mathbf{e}_z$, where $\mathbf{e}_x$, $\mathbf{e}_y$ and $\mathbf{e}_z$ are the unit vectors in the $x$--, $y$--, and $z$--directions, respectively. The component form of the governing equations in body-fitted coordinates is provided in Appendix \ref{app:A}.

Throughout this paper, we consider the weak curvature regime introduced by \cite{Dean_1928_channel}, in which $a^*$ is assumed to be $O(R^{2}d^*)$. Following convention, we introduce the Dean number, $De$, such that the nondimensional radius of curvature is given by 
\begin{eqnarray}
a=a^*/d^*=8R^2/De^2.\label{curvature}
\end{eqnarray}
Similar to Hall's analysis of G\"ortler vortices \citep{Hall_1983,Hall_1988}, we assume that $a$ varies slowly with $R^{-1}x$.



The channel walls are located at $y=\pm 1$, where the no-slip boundary conditions are imposed. In the $z$ direction, periodicity with period $2\pi/\beta$ is assumed, where $\beta$ denotes the spanwise wavenumber. To fully define the system, appropriate boundary conditions must also be specified at the upstream and downstream ends of the computational domain. Our interest lies in time-periodic solutions (i.e., global ECS) of this system. 

The need to specify upstream and downstream boundary conditions poses a difficulty not encountered in ECS computations with periodic domains. These conditions must be time periodic; otherwise, a time-periodic solution of the system cannot exist. In the scenario we consider, the curvature is assumed to be constant sufficiently far upstream of the inlet. This assumption allows us to use results obtained under periodic boundary conditions there. For example, when the curvature is zero upstream, the flow there reduces to plane Poiseuille flow, where travelling wave type ECS are well defined. In this setting, the inlet boundary condition naturally becomes time-periodic. As noted in section~\ref{sec:introduction}, generating travelling waves in experiments is not straightforward. Nevertheless, given the established importance of ECS in the dynamics of plane Poiseuille flow, this inlet condition is not unreasonable, at least as a simple test case.

Section~\ref{sec:2.2} briefly reviews the derivation of the BRE for readers who may not be familiar with the theory. This approximation enables streamwise integration of the governing equations, thereby eliminating the need to specify outlet boundary conditions. However, the BRE is limited to capturing fluid motions that evolve on the viscous time scale, $t^*=O(d^{*2}/\nu^*)$, and over the corresponding diffusive streamwise length scale, $x^*=O(U_b^*d^{*2}/\nu^*)$.

In contrast, self-sustained travelling waves in plane Poiseuille flow (i.e. local ECS at the inlet) typically exhibit dynamics on the convective spatio-temporal scale, $t=O(d^*/U_b^*)=O(R^{-1}d^{*2}/\nu^*)$ and $x=O(d^*)=O(R^{-1}U_b^*d^{*2}/\nu^*)$.
In section~\ref{sec:2.3}, we propose a computational approach that reconciles the slow scales captured by the BRE with the fast scales associated with the SSP-VWI. This section outlines the key modelling assumptions and explains the underlying ideas. The justification for these assumptions is provided in section~\ref{sec:2.4}.


\subsection{Derivation of BRE}\label{sec:2.2}

We take the Reynolds number $R$ as the large parameter in the asymptotic analysis. 
The rescaled variables suitable for the BRE are defined as $X=R^{-1}x$ and $T=R^{-1}t$. 
The effect of varying curvature can be incorporated by treating $De$ as a function of $X$. Substituting the asymptotic expansion
\begin{subequations}
\begin{eqnarray}
[u,v,w]=[U,R^{-1}V,R^{-1}W](X,y,z,T)+\cdots,\\
p=\Pi(X,T) +R^{-2}P(X,y,z,T)+\cdots,
\end{eqnarray}
\end{subequations}
into the governing equations (\ref{NSeq}) and retaining only the leading-order terms yields the BRE:
\begin{subequations}\label{eqBRE}
\begin{eqnarray}
[\partial_T+U\partial_X +V\partial_y +W\partial_z-\partial_y^2-\partial_z^2]
\left[ \begin{array}{c} U\\ V\\ W \end{array} \right]
+ \left[ \begin{array}{c} \Pi'\\ \partial_y P+De^2U^2/8 \\ \partial_z P  \end{array} \right] =0,~~~~~\\
\partial_XU+\partial_yV+\partial_zW=0.~~~~~
\end{eqnarray}
\end{subequations}
Recall that the channel walls are located at $y=\pm 1$.
This system closely resembles the Navier-Stokes equations for channel flow in Cartesian coordinates.

The effects of the wall curvature appear only in the term proportional to $De^2$. The origin of this term becomes apparent when considering the simplified case of constant curvature. In this case, cylindrical coordinates $(r,\varphi,z)$ can be used for the full governing equations, with the channel walls located at $r=a\pm 1$. The well-known Navier-Stokes equations in cylindrical coordinates, written in terms of $\mathbf{u}=u_r\mathbf{e}_r+u_{\varphi}\mathbf{e}_{\varphi}+u_z\mathbf{e}_z$, can be transformed into the previously defined curvilinear coordinate system by setting 
$u_r=-v, u_{\varphi}=u, u_z=w$, 
$y=a-r$, and $x=a\varphi$. 
In the radial momentum equation, 
the unsteady term $\partial_t u_r\approx R^{-2}\partial_T V$ balances with $r^{-1}u_{\varphi}^2\approx a^{-1}u^2\approx R^{-2}De^2U^2/8$. 

Note also that the advection operator $u_r\partial_r+r^{-1}u_{\varphi}\partial_{\varphi}+u_z\partial_z \approx U\partial_X +V\partial_y +W\partial_z$ is of formally size $O(R^{-1})$, and therefore balances with the viscous operator. However, second derivatives with respect to 
$X$ are negligible in the latter. Moreover, the pressure term in the azimuthal momentum equation can be approximated as $r^{-1}\partial_{\varphi}p\approx R^{-1}\Pi'$, and the term proportional to $\partial_XP$ is negligible. 
These reductions render the governing equations parabolic in form, treating 
$X$ as a time-like variable and allowing for downstream marching.

The case of non-constant curvature requires the use of the governing equations formulated in body-fitted coordinates, as presented in Appendix~\ref{app:A}. However, the underlying order-of-magnitude balance remains essentially unchanged. 


\subsection{Coupling BRE with fast scale local ECS}\label{sec:2.3}

Now, let us consider incorporating the effects of the fast-scale coherent structures into the BRE. Borrowing ideas from multiple-scale asymptotic analysis \citep[see][for example]{Howard_1977}, we introduce a fast scale variable $\theta=R\Theta(X,T)$ and decompose the velocity and pressure fields as
\begin{subequations}\label{decomp}
\begin{eqnarray}
\mathbf{u}=\overline{\mathbf{u}}(X,y,z,T)+\tilde{\mathbf{u}}(X,\theta,y,z,T),\label{veldecomp}\\
p=\overline{p}(X,y,z,T)+\tilde{p}(X,\theta,y,z,T).
\end{eqnarray}
\end{subequations}
The tilde components are assumed to be $2\pi$ periodic in $\theta$. In what follows, the overline also denotes averaging over $\theta$; thus the above decomposition represents a separation into the mean field and the fluctuation field. 
We will also decompose the governing equations into mean and fluctuation parts with respect to $\theta$.


That $\theta$ represents the fast scale becomes evident from the following relation 
\begin{eqnarray}
\frac{\partial}{\partial t}=-\Omega \frac{\partial}{\partial \theta} + R^{-1}\frac{\partial}{\partial T},\qquad
\frac{\partial}{\partial x}=\alpha \frac{\partial}{\partial \theta} + R^{-1}\frac{\partial}{\partial X},\label{multidel}
\end{eqnarray}
where we have written $\Omega=-\Theta_T$ and $\alpha=\Theta_X$. At this stage, the  `local' streamwise wavenumber $\alpha$ and frequency $\Omega$ of the wave-like motion on the fast scale may depend on both $X$ and $T$, provided they satisfy the consistency condition $\alpha_T+\Omega_X=0$.
If $\alpha$ is taken as a function of $X$ and $\Omega$ as a function of $T$, then the consistency condition is automatically satisfied, and we have
\begin{eqnarray}\label{phaseeq}
\Theta(X,T)=\int^X_0 \alpha(s)ds-\int^T_0 \Omega(\tau) d\tau.
\end{eqnarray}
The mean field is independent of the fast scale and therefore physically represents fluid motion that evolves slowly in the BRE spatio-temporal scales.


To reduce the governing equations and facilitate numerical computation, we introduce the following assumptions.
\begin{itemize}
\item Assumption (i): The governing equations (\ref{NSeq}) can be simplified to 
\begin{subequations}
\begin{eqnarray}\label{NSeq2}
(\partial_t +u\partial_x+v\partial_y+w\partial_z-R^{-1}(\partial_x^2+\partial_y^2+\partial_z^2))
\left[ \begin{array}{c} u\\ v\\ w \end{array} \right]+\left[ \begin{array}{c} \partial_x p\\ \partial_y p+u^2/a\\ \partial_z p \end{array} \right]=0,~~~~~~\\
\partial_xu+\partial_yv+\partial_zw=0.~~~~~~~~~
\end{eqnarray}
\end{subequations}
We substitute the decomposition (\ref{decomp}) into the governing equations and then separate them into mean and fluctuation parts by taking the $\theta$-average.
\item Assumption (ii): The mean equations can be parabolised by neglecting $\partial_X$ in the viscous and pressure terms.
\item Assumption (iii): 
Terms involving slow scale derivatives and curvature effects in the Reynolds stress terms can be omitted.
\item Assumption (iv): Likewise, terms involving slow scale derivatives and curvature effects in the fluctuation equations can also be neglected. 
\end{itemize}

\vspace{3mm}

We briefly explain here why they are expected to hold (see section \ref{sec:2.4} for justification using asymptotic analysis). Assumption (i) suggests that when the effect of curvature is weak, the flow may be approximated by the Cartesian Navier-Stokes equations, although the curvature terms appearing in the BRE must still be retained. Assumption (ii) has been confirmed for the BRE (section \ref{sec:2.2}). Assumptions (iii) and (iv) rely on the expectation that, when derivatives act on the fluctuation field, the contributions from the slow derivatives can be neglected.








Under the assumptions (i)-(iv), the mean part of the governing equations reduce to the form of the forced BRE upon writing
$\overline{\mathbf{u}}=[U,R^{-1}V,R^{-1}W]$, $\overline{p}=\Pi(X,T) +R^{-2}P(X,y,z,T)$:
\begin{subequations}\label{vortexeq}
\begin{eqnarray}
[\partial_T+U\partial_X +V\partial_y +W\partial_z-\partial_y^2-\partial_z^2]
\left[ \begin{array}{c} U\\ V\\ W \end{array} \right]
+ \left[ \begin{array}{c} \Pi'\\ \partial_y P+De^2U^2/8 \\ \partial_z P  \end{array} \right] =\mathbf{F},~~~~~~\label{BRE}\\
\partial_XU+\partial_yV+\partial_zW=0.~~~~~~~
\end{eqnarray}
This is not a surprising result, as in section~\ref{sec:3.1} we have checked the first two assumptions are valid for spatio-temporally slowly developing flow. 
When assumption (iii) holds, the forcing term can be compactly written as
\begin{eqnarray}
\mathbf{F}=-[R\overline{(\tilde{\mathbf{u}}\cdot \tilde{\nabla})\tilde{u}},R^2\overline{(\tilde{\mathbf{u}}\cdot \tilde{\nabla})\tilde{v}},R^2\overline{(\tilde{\mathbf{u}}\cdot \tilde{\nabla})\tilde{w}}],\label{Rstress}
\end{eqnarray}
\end{subequations}
which can be interpreted as the subgrid-scale Reynolds stress; here, $\tilde{\nabla}=[\alpha \partial_{\theta},\partial_y,\partial_z]$.
While the fluctuation part is obtained as
\begin{subequations}\label{waveeqq}
\begin{eqnarray}
(\overline{\mathbf{u}}\cdot \tilde{\nabla}-c\alpha  \partial_{\theta})\tilde{\mathbf{u}}+
(\tilde{\mathbf{u}}\cdot \tilde{\nabla})\overline{\mathbf{u}}
+(\tilde{\mathbf{u}}\cdot \tilde{\nabla})\tilde{\mathbf{u}}-\overline{(\tilde{\mathbf{u}}\cdot \tilde{\nabla})\tilde{\mathbf{u}}}
=-\tilde{\nabla} \tilde{p}+R^{-1}\tilde{\nabla}^2 \tilde{\mathbf{u}},\label{waveeq}\\
\tilde{\nabla}\cdot \tilde{\mathbf{u}}=0,\label{wave2}
\end{eqnarray}
\end{subequations}
writing $c(X,T)=\Omega(T)/\alpha(X)$. Thanks to assumption (iv), no derivatives with respect to $X$ and $T$ appear in the equations.

When $De = 0$, the travelling wave solutions of plane Poiseuille flow with phase speed $c$ exactly satisfy equations (\ref{vortexeq})--(\ref{waveeqq}). This follows from the fact that $\alpha$ becomes constant and we can use the fast-scale variable $\theta=\alpha (x-ct)$.
The travelling waves fit within the periodic box $[0,2\pi/\alpha] \times [-1,1] \times [0,2\pi/\beta]$ in the $(x,y,z)$ coordinates, and therefore the dependence on the slow variables $X$ and $T$ disappears from the global flow. Even if $De$ is a nonzero constant, travelling-wave-type ECS solutions persist in the reduced system, though only for the approximate system (\ref{NSeq2}) and not for the full cylindrical coordinate system.


The reduced equations (\ref{vortexeq})--(\ref{waveeqq}) can thus be viewed as a means of incorporating Reynolds stresses generated by the SSP-VWI into the BRE via ECS. Alternatively, the formulation may be regarded as a generalisation of ECS studies in periodic domains that captures the spatial evolution of the system. Importantly, this is not merely a local parallel approximation; rather, it naturally accounts for what are referred to as non-parallel effects in boundary layer theory.



\subsection{Justification of the assumptions}\label{sec:2.4}

The assumptions used to derive the reduced equations can be justified using a large Reynolds number asymptotic analysis. Specifically, our argument builds on the vortex/Rayleigh-wave interaction theory developed by \cite{hall1991}, in which a multiple-scale asymptotic analysis is applied to boundary-layer flows. In this theory, the mean field and the fluctuation field are referred to as the vortex component and the wave component, respectively. The former component is assumed to follow an asymptotic expansion similar to that in the BRE
\begin{subequations}\label{vortexVWIexp}
\begin{eqnarray}
\overline{\mathbf{u}}=[U,R^{-1}V,R^{-1}W](X,y,z,T)+\cdots,\\
\overline{p}=\Pi(X,T) +R^{-2}P(X,y,z,T)+\cdots,
\end{eqnarray}
\end{subequations}
while the latter component is expanded as
\begin{subequations}\label{waveVWIexp}
\begin{eqnarray}
\tilde{\mathbf{u}}=\delta^{1/2} R^{-1}e^{i\theta}\hat{\mathbf{u}}(X,y,z,T)+\text{c.c.}+\cdots,\\
\tilde{p}=\delta^{1/2} R^{-1}e^{i\theta}\hat{p}(X,y,z,T)+\text{c.c.}+\cdots,
\end{eqnarray}
\end{subequations}
where $\delta=R^{-1/3}$ and c.c. stands for complex conjugate. 

Since the wave part is predominantly advected by $U\partial_x$, both the viscous and curvature terms can be neglected. Furthermore,  the small amplitude of the wave justifies linearisation. In fact, the leading order part of the fluctuation equations can be obtained as 
\begin{subequations}\label{Rayleigh}
\begin{eqnarray}
i\alpha(U-c)\hat{u}+\hat{v}\partial_y U+\hat{w}\partial_z U=-i\alpha \hat{p},\\
i\alpha(U-c)\hat{v}=-\partial_y \hat{p},\\
i\alpha(U-c)\hat{w}=-\partial_z \hat{p},\\
i\alpha \hat{u}+\partial_y\hat{v}+\partial_z\hat{w}=0.
\end{eqnarray}
\end{subequations}
These equations can be reduced to a single equation for $\hat{p}$, namely the Rayleigh equation, which becomes singular when $U=c$. This singularity at the critical level $y=f(z,X,T)$ must be regularised by viscous effects within a critical layer of thickness $\delta=R^{-1/3}$.
Near the critical level, the wave velocity components $\hat{u}, \hat{v}, \hat{w}$ behave like $1/(U - c)$, and therefore the wave amplitude increases inversely with the distance from the critical level. As a result, it is amplified by a factor of $O(\delta^{-1})$ within the critical layer. 

Recalling in the discussion in sections~\ref{sec:2.2} and \ref{sec:2.3}, it is evident that the mean part of the governing equations reduces to the forced BRE, and the calculation of the Reynolds stress (\ref{Rstress}) must take into account the singular behaviour of the wave. The size of $\mathbf{F}$ can be estimated as $R^2$ times the square of the wave amplitude. Outside the critical layer, this amplitude is $O(\delta^{1/2}R^{-1})$, so the Reynolds stress is negligible. 
However, the integration of $\mathbf{F}$ across the critical layer becomes $R^2\times (\delta^{-1/2}R^{-1})^2\times \delta=O(1)$, and therefore, the feedback from the wave to the vortex is not negligible.

Note that to carry out a rigorous asymptotic analysis, the method of matched asymptotic expansions must be employed. The behaviour of the solution of (\ref{Rayleigh}) near the critical level can be analysed using the method of Frobenius expansion (see \cite{Deguchi_2019}, for example). 
To match them with the flow within the critical layer, one must employ asymptotic expansions different from those used in the outer region. 
After lengthy algebraic manipulations, we can show that the concentrated Reynolds stress within the critical layer leads to jumps in the derivatives of $V$, $W$, and in the pressure 
$P$. These jump conditions, together with the BRE and the Rayleigh equation, form a closed system that does not involve $R$, as shown in \cite{hall1991} and \cite{HALL_SHERWIN_2010}.

Notably, \cite{HALL_SHERWIN_2010} found that the behaviour of the ECS computed by \cite{wang2007} for plane Couette flow is accurately captured by the vortex/Rayleigh-wave interaction theory, with $U$, $(V,W)$, and $(\hat{u},\hat{v},\hat{w})$ corresponding to the streak, roll, and wave components, respectively, of the self-sustaining process \citep{waleffe1997}.
Similar observations have been repeatedly confirmed in parallel flows; see \cite{deguchi2014}, \cite{Brand_Gibson_2014} and \cite{Deguchi_Hall_2016}. 
\cite{hall1991} also considered vortex/Tollmien-Schlichting wave interaction (see \cite{hall_smith_1988} also). A validation of this theory using ECS was later carried out by \cite{Dempsey_Deguchi_Hall_Walton_2016} using plane Poiseuille flow. In both versions of the vortex–wave interaction theory, the fluctuation part of the equations is linear in the wave component, and therefore the neutral mode is proportional to a single Fourier mode (see (\ref{waveVWIexp})). 

Our reduced system (\ref{vortexeq})--(\ref{waveeqq}) retains more terms than the vortex–wave interaction theory. One advantage of this approach is that it allows an ECS to be directly used as the inlet condition. Another benefit is that it can accommodate wave fields involving multiple Fourier modes. An example of such a case appears in the strongly nonlinear critical layer theory by \cite{Smith_Bodonyi_1982} (for validation using ECS, see \cite{Deguchi_Walton_2018}).
Also, when coherent structures appear in the free stream of boundary layers \citep{deguchi2014} or at the Kolmogorov scale \citep{Deguchi_2015}, all terms in the Navier–Stokes equations can become important at the short scale. Even in such extreme cases, assumptions (i)–(iv) remain valid, and our reduced equations can still be applied.

We also wish to comment on the differences between our method and another major spatial marching approach, the parabolised stability equations \citep[PSE, see the seminal review by][]{Herbert_1997}. PSE has been widely used to efficiently identify transition points to turbulence in various aerodynamic design problems \citep{Chang2004}.
Rather than using (\ref{waveeqq}) for the fast scale, PSE adopts a different equation that is not compatible with using an ECS as the inlet condition. A more fundamental difference lies in the treatment of the streamwise wavenumber $\alpha$, which in PSE is allowed to be complex and is updated at each step of the spatial marching according to a prescribed rule. The imaginary part of $\alpha$ represents the spatial growth rate of the fluctuation field. Consequently, if this growth rate is nonzero, we cannot mathematically exclude the possibility that the Reynolds stresses vary on the fast scale. This violates a key assumption of multiple scale analysis, namely that the mean field depends only on the slow spatial variables.
Since the Reynolds stress is a nonlinear quantity, this issue is intrinsic when seeking finite-amplitude perturbations.


Although reasonably good agreement between PSE results and asymptotic theory has often been observed, this agreement breaks down over long spatio-temporal scales. Readers interested in asymptotic analyses of spatially developing flows may consult \cite{Wu_2019}. It should also be noted that the objectives of PSE and our approach differ fundamentally. While the former seeks to capture the evolution of the flow field near the transition point, the latter aims to uncover the key ECS that organise the phase space.

\begin{figure}
\centering
\begin{overpic}[width=0.85 \textwidth]{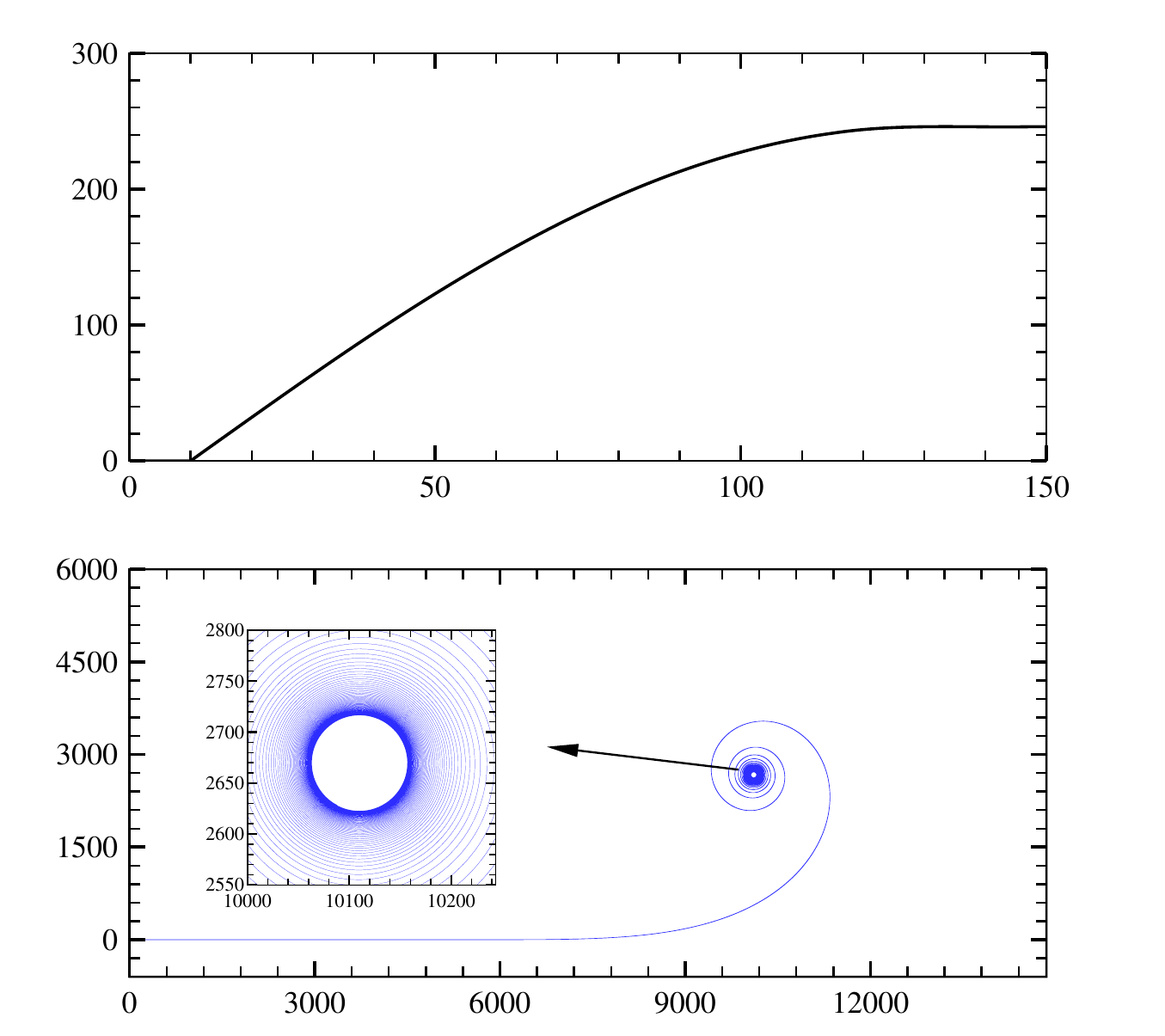}
\put(-3,83){(a)}
\put(-3,39.3){(b)}
\put(0,22){\begin{turn}{90}{$\check{y}$}\end{turn}}
\put(0,70){\begin{turn}{90}{$De$}\end{turn}}
\put(50,-1){{$\check{x}$}}
\put(50,44){{$X$}}
\end{overpic}
\caption{
Channel curvature. (a) Dean number profile defined in section \ref{sec:3.1}, used for spatial marching in sections~\ref{sec:4.1} and~\ref{sec:4.2}. (b) Channel centreline plotted in Cartesian coordinates.
}
\label{fig:Rp_zAm}
\end{figure}
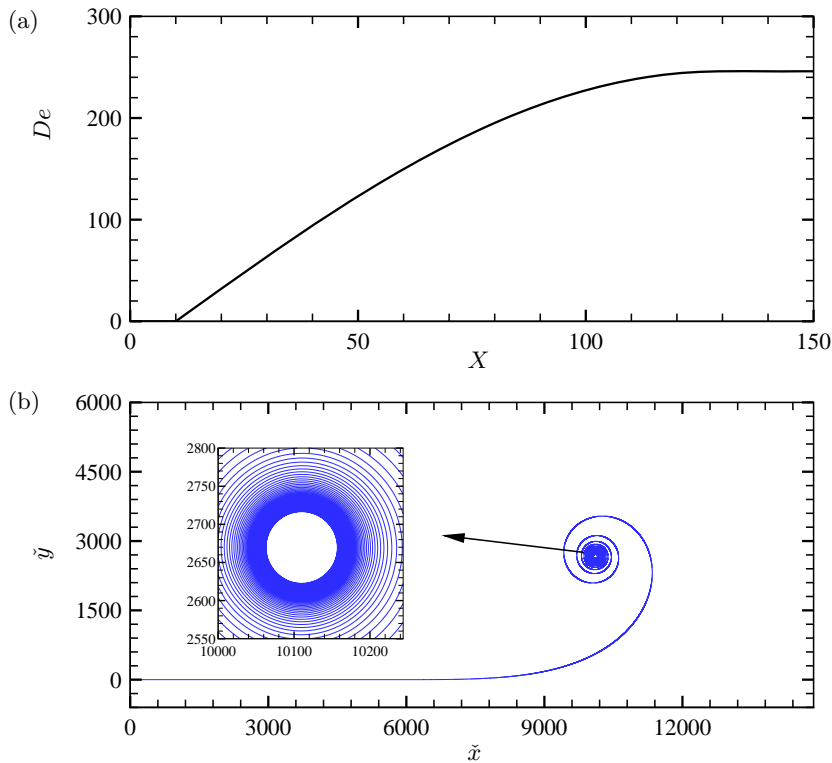

\section{Numerical method}\label{sec:3}

\subsection{Computational setup}\label{sec:3.1}
In section~\ref{sec:4}, we numerically investigate a curved channel 
with the Dean number $De(X)=De_{\text{max}}\,\chi(X)$, where
\begin{equation}
 \chi(X) =
\begin{cases}
  0, &  0 \leq X < X_A, \\
  [\frac{1}{2}-\frac{1}{2}\cos(\frac{X-X_A}{X_B-X_A}\pi)]^{\frac{1}{2}} ,  &  X_A \leq X < X_B, \\
  1,&  X_B \leq X.
\end{cases}\label{Deprof}
\end{equation}
Our choice of the parameters are $De_{\text{max}}=\sqrt{60500}\approx 245.9$,  $X_A=10$ and $X_B=130$. The profile $De(X)$ is shown in figure \ref{fig:Rp_zAm}-(a).
According to the stability analysis of Dean flow in a uniformly curved channel, an instability arises when $De$ exceeds 35.9 \citep[see][for example]{GIBSON_COOK_1974,mizushima1998}.
Thus, $De_{\text{max}}$ used in our study is large enough to induce Dean vortices. The flow in the region $X < 0$ is assumed to be plane Poiseuille flow, where an edge state is assumed to be sustained. 

We choose $R = 550$ and carry out spatial marching of the reduced equations (\ref{vortexeq})--(\ref{waveeqq}) over the interval $X \in [0, 150]$. 
To visualise the actual shape of the channel, we introduce Cartesian coordinates $(\check{x}, \check{y}, \check{z})$, which coincide with the curvilinear coordinates $(x, y, z)$ at the origin (figure \ref{fig:sketch}). 
Figure~\ref{fig:Rp_zAm}-(b) shows the channel centreline reconstructed from the radius of curvature (\ref{curvature}).
The channel forms a spiral with a total arclength of $550\times 150=82500$. Computing ECS over such a large domain using the full Navier–Stokes equations would be impractical. Towards the end of the spiral, the radius of curvature reduces to 40, which nevertheless remains much larger than the channel width.

Since $De(X)$ vanishes at the inlet $X=0$, a travelling-wave-type ECS of plane Poiseuille flow can naturally serve as the initial condition for spatial marching. Note that such a travelling wave is independent of the slow time variable $T$, and therefore the global solution of the reduced system also becomes independent of $T$. The fast scale variable simplifies
\begin{eqnarray}\label{phaseeq2}
\theta=R\int^X_0 \alpha(s)ds- \Omega t,
\end{eqnarray}
where $\Omega$ is a constant.
The solution constructed by spatial marching is a time-periodic solution with period $2\pi/\Omega$. The marching problem depends on the frequency $\Omega$ and the spanwise wavenumber $\beta$; these must be determined so that they are consistent with the inlet ECS.
The pressure gradient $\Pi'(X)$ and the local wavenumber $\alpha(X)$ are determined as part of the solution.


\subsection{Inlet conditions: ECS of plane Poiseuille flow} \label{sec:3.2}

\begin{figure}
\centering
\begin{overpic}[width=0.95 \textwidth]{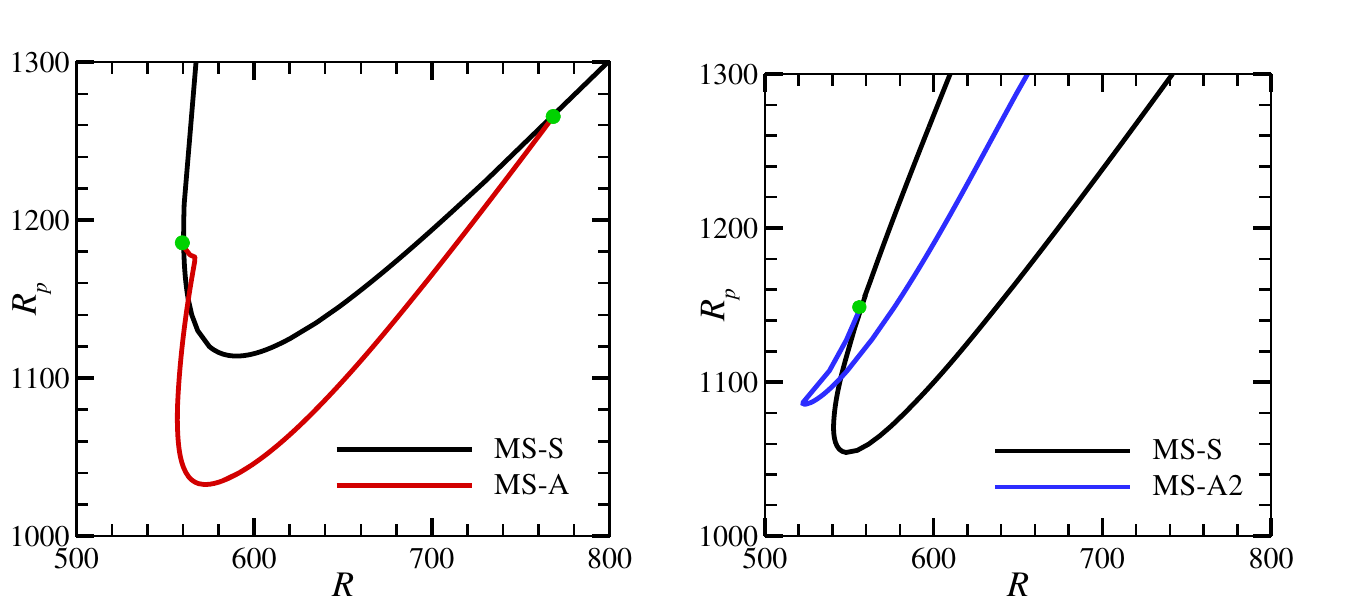}
\put(-3,39.3){(a)}
\put(47,39.3){(b)}
\end{overpic}
\caption{Symmetry-breaking bifurcation from the MS-S solution
branch in plane Poiseuille flow.
(a) $(\alpha,\beta)=(1.0,2.0)$; (b) $(\alpha,\beta)=(1.01,2.72)$. 
}
\label{fig:bd_MS-MS-A2}
\end{figure}

\begin{figure}
\centering
\begin{overpic}[width=0.8 \textwidth]{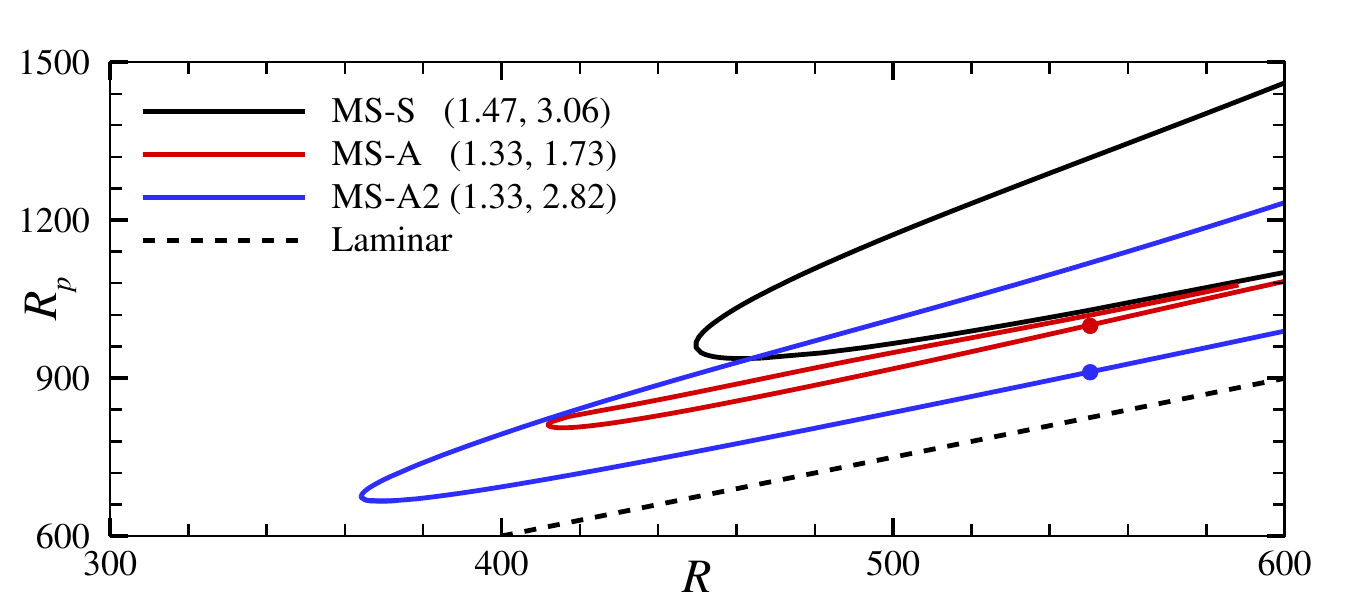}
\end{overpic}
\caption{Bifurcation diagram for plane Poiseuille flow, showing the pressure-based Reynolds number $R_p$ versus the bulk Reynolds number $R$. The wavenumbers are optimised to minimise the value of $R_p$ for each solution.
The solid lines represent the ECS. The solution names are provided in the legend, along with the corresponding wavenumbers $(\alpha, \beta)$.
The red and blue circles indicate the initial solutions used in the spatial marching shown in figures \ref{fig:ND13_vis} and \ref{fig:new_vis}, respectively (see table~\ref{table:inlet} also).
}
\label{fig:bd_R}
\end{figure}

\begin{figure}
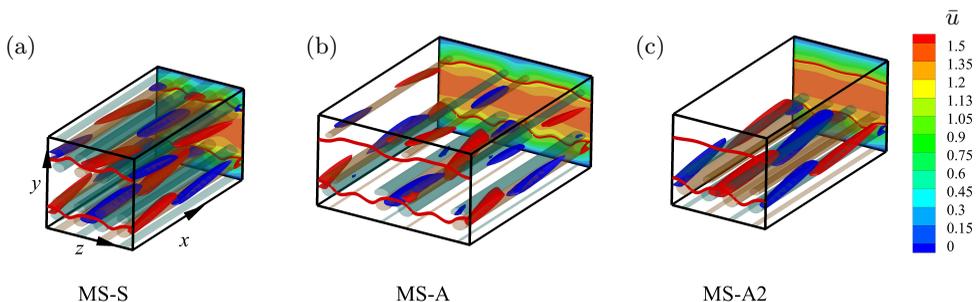

\centering
\begin{overpic}[width=0.95 \textwidth]{A/fig5/fig5.pdf}
\put(-1,32){(a)}
\put(30,32){(b)}
\put(64,32){(c)}
\put(96,35){{$\bar u$}}
\end{overpic}
\caption{Flow visualisation of the lower branch ECS in plane Poiseuille flow at $R=550$. 
The optimal wavenumbers used in figure \ref{fig:bd_R} are adopted here.
The red/blue surfaces depict 50\% maximum/minimum
value of $\partial_y\tilde{w} - \partial_z\tilde{v}$.
The brown/green surfaces are 50\% maximum/minimum value of $\partial_y\overline{w}-\partial_z \overline{v}$.
The colourmap illustrates $\overline{u}$. The thick red lines indicate the location of the critical levels, where $\overline{u}=c$. 
In the absence of slow-scale dependence, $(\overline{u},\overline{v},\overline{w})$ correspond to the streamwise averaged velocity fields.
}
\label{fig:fig5}
\end{figure}

We employ two types of travelling wave solutions of plane Poiseuille flow, both obtained by solving the full Navier–Stokes equations in a periodic box $(x,y,z)\in [0,2\pi/\alpha]\times [-1,1]\times [0,2\pi/\beta]$ using the Newton–Raphson method. 
Both solutions are found via symmetry breaking bifurcations from the MS-S solution reported by \citet{Nagata_Deguchi_2013}.
To describe the symmetry of the solution, we use the fact that a travelling wave with phase speed 
$c$ is a function of $\theta$, $y$, and $z$, where
\begin{eqnarray}
\theta=\alpha (x-ct).
\end{eqnarray}
The MS-S solution has the three symmetries
: the mirror symmetry (MS) in the spanwise direction
\begin{eqnarray}
~[u,v,w](\theta,y,z)=[u,v,-w](\theta,y,-z+\pi/\beta),
\end{eqnarray}
the shift-reflection symmetry, 
\begin{eqnarray}
~[u,v,w](\theta,y,z)=[u,v,-w](\theta+\pi,y,-z), 
\end{eqnarray}
and top-bottom reflectional symmetry
\begin{eqnarray}
~[u,v,w](\theta,y,z)=[u,-v,w](\theta,-y,z).\label{wallsym}
\end{eqnarray}
Figures \ref{fig:bd_MS-MS-A2}-(a) and (b) illustrate the bifurcation diagram for two sets of wavenumber pairs $(\alpha,\beta)$. The vertical axis is $R_p=U_c^*d^*/\nu$ defined using the
centreline velocity of laminar flow, $U_c^*$; this Reynolds number is proportional to the pressure gradient required to maintain the bulk velocity at unity. At the bifurcation point indicated by the green circle, the symmetry of MS-S in the wall-normal direction (\ref{wallsym}) is broken, while the other two symmetries are preserved. We refer to the solutions shown in panels (a) and (b) as MS-A and MS-A2, respectively (thus S and A represent top–bottom symmetry and asymmetry, respectively). 


In figure \ref{fig:bd_R} the wavenumber pairs $(\alpha,\beta)$ are selected to minimise the onset $R_p$ of each solution. This bifurcation diagram, presented in nearly the same format as figure 5 of \cite{Nagata_Deguchi_2013}, clearly demonstrates that MS-A and MS-A2 are distinct solutions, despite sharing the same flow symmetry. As the Reynolds number increases, two solution branches emerge abruptly via a saddle node bifurcation.
For the inlet conditions of the spatial marching problem presented in section~\ref{sec:4}, we use the lower-branch solutions of MS-A and MS-A2 at $Re = 550$ (indicated by circles). The parameters of these solutions are summarised in table~\ref{table:inlet}, and the corresponding flow fields are shown in figure \ref{fig:fig5}-(b) and (c).
Note that the frequency $\Omega$ of the global problem must be equal to the product $\alpha c$ determined from the inlet ECS. Also, the pressure gradient at $X=0$ is given by $\Pi'=-2R_p/R$.

\begin{table}
	\centering
	\begin{tabular}{cccccccc}
		   Inlet ECS   & $\alpha$   & $\beta$ & $R$ & $c$& $R_p$ & $\Omega$ & $\Pi'$\\
		\hline
        MS-A &1.33&1.73&550&1.08&999.21&1.44&$-3.63$\\
MS-A2 & 1.33&2.82&550&1.14&911.06&1.52&$-3.31$
	\end{tabular}
	\caption{Parameters and key physical properties of the initial plane Poiseuille ECS at the inlet, used in the spatial marching problem discussed in section 4. These solutions correspond to the circles in figure \ref{fig:bd_R}.}
 \label{table:inlet}
\end{table}

It is known that the lower branch solutions are well described by the vortex/Rayleigh-wave interaction theory, even at moderately high Reynolds numbers \citep{HALL_SHERWIN_2010,deguchi2014,Brand_Gibson_2014}. 
Indeed, the flows shown in figure \ref{fig:fig5} exhibit all three elements of the SSP-VWI. 
The presence of waves and rolls can be identified from the isosurfaces of streamwise vorticity in the fluctuation and mean fields, respectively.
The streak field can be seen in the colourmap.
The two red curves indicate the critical levels, around which a strong wave field is generated, as predicted by \cite{HALL_SHERWIN_2010}.

Another important feature of the lower-branch solution is that, when the Navier–Stokes equations are viewed as a dynamical system, it lies on the basin boundary separating the laminar and turbulent attractors in the phase space \citep{Itano_Toh_2001,wang2007,Eckhardt2007,Deguchi_Hall_2016}. Our inlet conditions correspond to an edge state in the sense that small perturbations around them can lead either to turbulence or relaminarisation.
(Note that some studies define edge states more narrowly, referring specifically to the converged solution obtained through the flow control algorithm known as edge tracking.)





\subsection{Numerical scheme}

To solve (\ref{vortexeq})--(\ref{waveeqq}) numerically, we write the mean field and the fluctuation field as
\begin{eqnarray}
\left[ \begin{array}{c} U\\ V\\ W \end{array} \right]=
\left[ \begin{array}{c} -\Psi_z+\Xi_y \\ -\Phi_{zz}-\Xi_X\\ \Phi_{yz}+ \Psi_{X}\end{array} \right],\qquad
\left[ \begin{array}{c} \tilde{u}\\ \tilde{v}\\ \tilde{w} \end{array} \right]=
\left[ \begin{array}{c} \alpha \phi_{\theta y}-\psi_z \\ -\alpha^2\phi_{\theta \theta}-\phi_{zz}\\ \phi_{yz}+\alpha \psi_{\theta} \end{array} \right],\label{potentials}
\end{eqnarray}
where partial derivatives have been expressed using subscripts to simplify the notation. As a result, the primitive variables $[U, V, W, P, \tilde{u}, \tilde{v}, \tilde{w}, \tilde{p}]$ can be reduced to a smaller set of variables $[\Xi, \Phi, \Psi,\phi, \psi]$. 

Here, $\phi(X,\theta,y,z)$ and $\psi(X,\theta,y,z)$ represent the poloidal and toroidal potentials of the fluctuation field, respectively. The use of such potentials is well established in computations with periodic boxes (see \cite{McBain_2005} for example), and it naturally extends to the fluctuation field in our formulation. The governing equations for $\phi$ and $\psi$ can be found by applying the operators $\mathbf{e}_y \cdot \tilde{\nabla} \times \tilde{\nabla} \times$ and $\mathbf{e}_y \cdot \tilde{\nabla} \times$ to equation (\ref{waveeq}), respectively. The continuity is automatically satisfied in this formulation.

A similar potential decomposition can also be applied to the mean field. However, for the BRE, the $z$-averaged components require special treatment, as discussed by \cite{Deguchi_Hall_Walton_2013}.
Thus, we decompose the field into its $z$-average and the reminder, and introduce the potentials 
$\Phi(X,y,z)$ and $\Psi(X,y,z)$ for the latter, ensuring they have zero $z$-average. Then, for the components $\langle U \rangle$ and $\langle V \rangle$, where angle brackets denote averaging in the $z$ direction, we introduce a streamfunction $\Xi(X,y)$. 
If the inlet condition satisfies mirror symmetry in the 
$z$-direction, the global solution inherits this symmetry, and thus $\langle W \rangle=0$. 
%
The pressure $P$ can be eliminated from the $z$ fluctuation part of (\ref{BRE}), yielding two equations for $\Phi$ and $\Psi$. The equation for $\Xi$ is obtained by taking the $z$-average of the streamwise component of (\ref{BRE}):
\begin{eqnarray}
\partial_X\langle U^2\rangle+\partial_y\langle UV\rangle-\partial_y^2 \langle U\rangle+\Pi'=\langle\mathbf{e}_x\cdot \mathbf{F}\rangle.\label{meaneq}
\end{eqnarray}
The three equations are equivalent to the forced BRE  and can be integrated in $X$ by applying finite differences.

If the fluctuation field is known, the problem reduces to solving the BRE with extra forcing terms. Conversely, if the mean field is known, the equations for $\phi$ and $\psi$ can be solved by the Newton–Raphson method, in much the same way as in periodic box ECS computations. Thus, the most straightforward strategy for solving (\ref{vortexeq})--(\ref{waveeqq}) is to iteratively update the mean and fluctuation fields at each spatial marching step until convergence is achieved. However, unfortunately convergence is not guaranteed in general. Therefore, in this study, we solve both fields simultaneously to ensure robust convergence. Upon applying backward differences to the $X$ derivatives, we obtain implicit inhomogeneous equations for the updated fields. This system can be efficiently solved using the Newton–Raphson method, by extending the code developed by \cite{Deguchi_Hall_Walton_2013}.

The Fourier spectral method is used in the $\theta$ and $z$ directions, while the Chebyshev collocation method is employed in the $y$ direction. 
We expand $\Xi$, $\Phi$, and $\phi$ in the basis $(1 - y^2)^2 T_l(y)$, and $\Psi$ and $\psi$ in the basis $(1 - y^2) T_l(y)$ so that 
the no-slip conditions
\begin{eqnarray}
\Xi=\partial_y\Xi=\Phi=\partial_y\Phi=\phi=\partial_y\phi=\Psi=\psi=0\qquad \text{at} \qquad y=\pm 1
\end{eqnarray}
are fulfilled.
Here, $T_l(y)$ is the $l$-th Chebyshev polynomial, and the collocation points are $y_k=\cos (\frac{k+1}{K+2}\pi), k=0,\dots,K$.

The choice of basis functions for $\Xi$ imposes four conditions that ensure $\langle U\rangle=\langle V\rangle=0$ on both walls. Thus integrating continuity $\langle U\rangle_X+\langle V\rangle_y=0$ with respect to $y$ over $[-1,1]$, we see that the flux $\int^1_{-1}\langle U\rangle dy$ is unchanged in $X$. 
To obtain a fourth-order differential equation for $\Xi$, we use the equation obtained by differentiating (\ref{meaneq}) with respect to $y$. 
The eliminated $\Pi'$ can be recovered using $\Pi' = (\partial_y^2 \langle U \rangle) |_{y = -1}$, which follows from evaluating (\ref{meaneq}) at the lower wall.
It can also be computed by integrating (\ref{meaneq}) in $y$:
\begin{eqnarray}
\partial_X\int^1_{-1}\langle U^2\rangle dy-[\partial_y \langle U\rangle]^1_{-1}+2\Pi'=0.
\end{eqnarray}

We analytically compute the Jacobian matrix and solve the linear system arising in each Newton step using a direct method.
To eliminate the translational degree of freedom in the $\theta$ direction, we set the imaginary part of one spectral coefficient associated with the $e^{i\theta}$ mode to zero. 
The converged solution of the Newton iterations is obtained by treating $c$ as an unknown while keeping $\alpha$ fixed. To match $\alpha c$ with the given $\Omega$, a line search is used to find the optimal $c(\alpha)$ that minimises $|1-\frac{\alpha c}{\Omega}|$ to $O(10^{-7})$. The convergence of the search is fast, so each spatial marching step incurs a cost comparable to that of computing an ECS in a small periodic domain.

In the absence of the fluctuation field, the code is validated against the nonlinear G\"ortler vortex solution obtained by \cite{Hall_1988}. We also verified that, when 
$De=0$, the solution for plane Poiseuille flow presented in section~\ref{sec:3.2} can be reproduced.

\begin{figure}
\centering
\begin{overpic}[width=0.95 \textwidth]{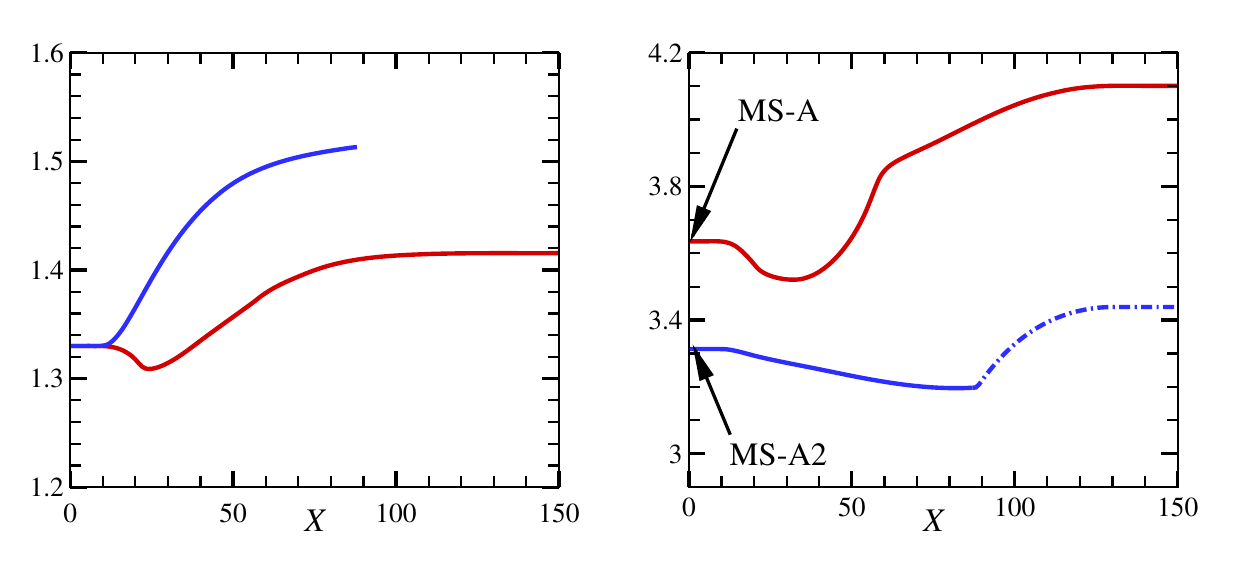}
\put(-3,39.3){(a)}
\put(47,39.3){(b)}
\put(50,22){\begin{turn}{90}{$-\Pi'$}\end{turn}}
\put(0,22){\begin{turn}{90}{$\alpha$}\end{turn}}
\end{overpic}
\caption{
Key quantities obtained from the spatial marching problem in the curved channel. 
The red and blue curves represent the cases where MS-A and MS-A2 in table 1 are used as the inlet conditions, respectively.
(a) Local streamwise wavenumber $\alpha$.
At $X=88.35$, the blue curve terminates because the fast-scale dependence disappears and the Dean vortex solution emerges. 
(b) Pressure gradient $-\Pi'$.
The dot-dashed curve corresponds to the Dean vortex solution.
}
\label{fig:fig6}
\end{figure}
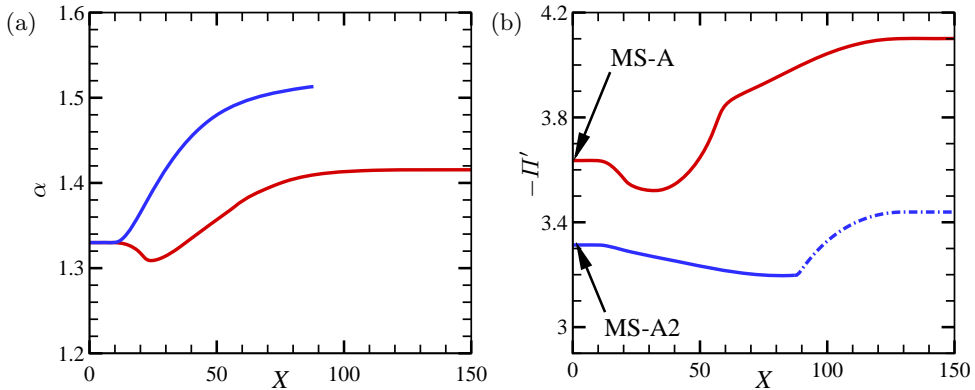

\section{Numerical results}\label{sec:4}

\subsection{Marching with MS-A as the inlet condition}\label{sec:4.1}

We begin by using MS-A from table~\ref{table:inlet} as the inlet condition and describe the typical outcome of the spatial marching. As noted in section~\ref{sec:3.1}, both $\alpha$ and $\Pi'$ are obtained as part of the solution. These results are shown as the red curves in figure \ref{fig:fig6}.

The channel curvature profile, given in (\ref{Deprof}), is designed to include a straight section over the range $X \in (0,10)$. In this region, the curvature is zero, and the quantities $\alpha$ and $\Pi'$, plotted in figure \ref{fig:fig6}, remain constant. This confirms that MS-A is stable under spatial marching in the plane Poiseuille flow, despite being unstable in direct numerical simulations.
As the curvature gradually increases along the section $X \in [10,130)$, both $\alpha$ and $\Pi'$ undergo non-trivial changes. Once the curvature becomes constant again beyond $X = 130$, these quantities rapidly converge to steady values.

\begin{figure}
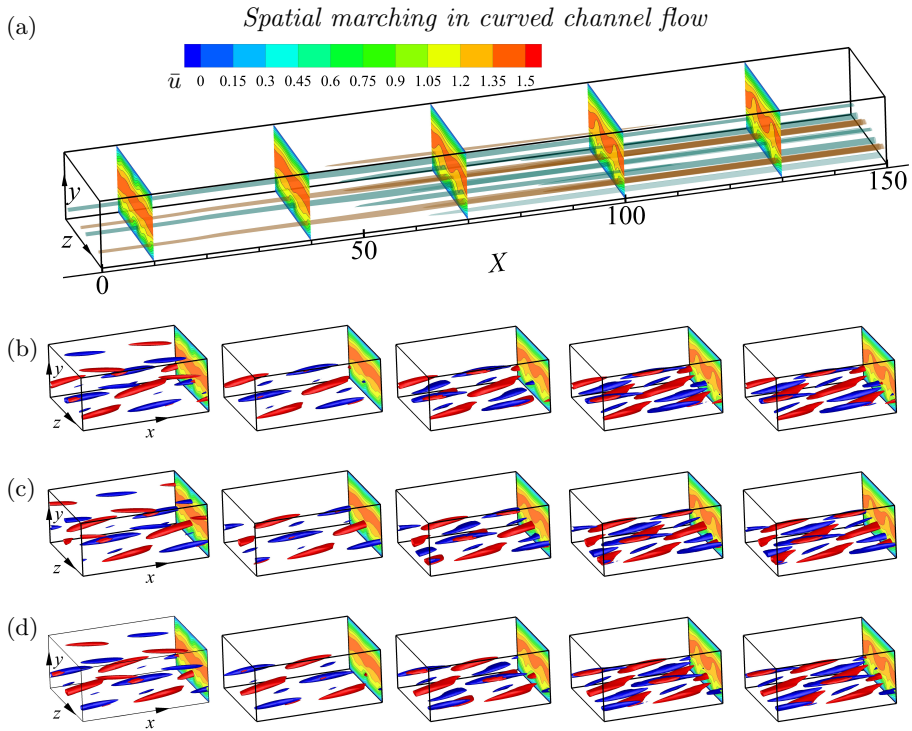

\centering
\begin{overpic}[width=0.85 \textwidth]{A/fig7/fig7.pdf}
\put(-4,80){(a)}
\put(-4,43){(b)}
\put(-4,27){(c)}
\put(-4,11){(d)}
\put(14.6,73.8){{$\bar u$}}
\end{overpic}
\caption{The flow fields obtained for the curved channel flow using the spatial marching method. The channel curvature is defined by $De(X)$ with $De_{\text{max}}=\sqrt{60500}\approx 245.9$, as shown in figure \ref{fig:Rp_zAm}. 
The inlet condition at $X=0$ is the lower branch of MS-A (see table~\ref{table:inlet}). (a) Mean field. Isosurfaces show the roll component, and the colourmap illustrates the streak component. (b) Snapshots of the fluctuation field at the locations corresponding to the colourmap positions in panel (a). Panels (c) and (d) show the same plots as (b), taken at one-quarter and one-half of a time period later, respectively. See the caption of figure \ref{fig:fig5} for precise definition of the isosurfaces.
%
}
\label{fig:ND13_vis}
\end{figure}

The numerical computation also determines the mean field, $\overline{\mathbf{u}}$, and the fluctuation field, $\tilde{\mathbf{u}}$, as defined in (\ref{veldecomp}), at each streamwise location $X$.
Since their spatial scales differ significantly, we choose to plot them separately in figure~\ref{fig:ND13_vis}. Plotting the mean (slow scale) field in the $(X,y,z)$ coordinates is straightforward, as shown in panel (a). The isosurfaces represent the streamwise vorticity, $\partial_y\overline{w} - \partial_z\overline{v}$, while the colourmap shows the streamwise velocity, $\overline{u}$. These quantities are steady and correspond to the roll and streak components of the SSP-VWI, respectively.
In contrast, plotting the fluctuation (fast scale) field requires computing $\theta$ using (\ref{phaseeq2}). For instance, when fixing $t = 0$, the value of $\theta$ at each $X$ is determined by integrating the streamwise wavenumber $\alpha$ shown in figure~\ref{fig:fig6}-(a). 
Figure \ref{fig:ND13_vis} (b)-(d) shows the snapshots of the streamwise vorticity, $\partial_y\tilde{w} - \partial_z\tilde{v}$; they visualise the wave component in the SSP-VWI. 
Since the fluctuation field has a short characteristic streamwise length scale of $x=O(1)$, we use the small boxes defined around $X=X_0=10, 40, 70, 100$ and $130$. The boxes have dimensions $[0,2\pi/\alpha(X_0)] \times [-1,1] \times [0,2\pi/\beta]$ in the shifted original coordinates $(x-RX_0,y,z)$. 


The fluctuation fields in panels (b–d) appear nearly as travelling waves, but not exactly. This is because both the fluctuation field functions $\tilde{\mathbf{u}}$ and the phase $\theta$ vary with the streamwise position.
Note also that the boxes in panels (b)-(d) maintain the correct aspect ratio in the original coordinates $(x,y,z)$. When converted to the 
$X$-coordinate, the local wavelength of the fluctuation wave is only 
$O(0.01)$. 


Figure~\ref{fig:ND13_vis} shows that the centrifugal force induced by the channel curvature reinforces the mean field in the lower half of the channel, consistent with the findings of \cite{Dean_1928_channel} and subsequent studies. However, there is no sharp boundary between the self-sustained coherent structures and the Dean vortices, as the fluctuation part of the solution persists throughout the spatial marching. At the inlet, wave-like structures associated with the fluctuation field are present near both walls, but those in the upper half of the channel are gradually suppressed. This suppression occurs because the weakening of the roll–streak structure by the centrifugal force disrupts the SSP-VWI cycle.





\begin{figure}
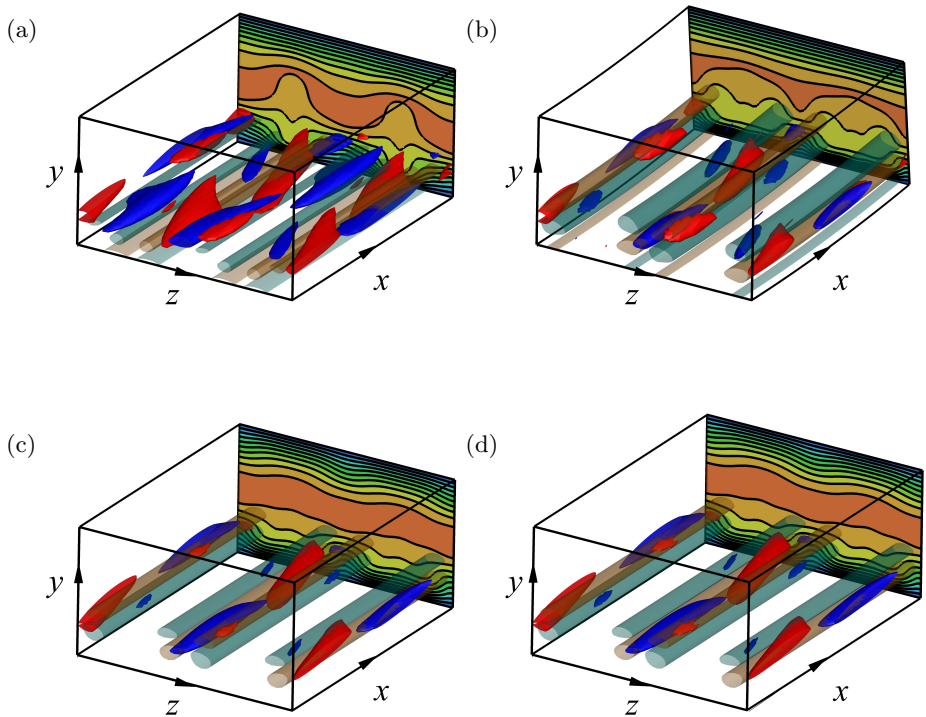

\centering
\begin{overpic}[width=0.9 \textwidth]{A/fig8/fig8_new_2.pdf}
\put(0,80){(a)}
\put(50,80){(b)}
\put(0,35){(c)}
\put(50,35){(d)}
\end{overpic}
\caption{ Comparison of constant curvature channel flow solutions obtained from the reduced and full systems. (a) Solution obtained from the spatially marching computation shown in figure \ref{fig:ND13_vis}.
This solution is taken near the outlet at 
$X=150$ and is plotted in a local small box, similar to figure \ref{fig:ND13_vis}-(b).
See the caption of figure \ref{fig:fig5} for precise definitions of the isosurfaces and colourmaps.
(b) The corresponding travelling wave solution of the Navier–Stokes equations in the annulus, with the bulk Reynolds number, frequency, and radius of curvature matching those of (a) (i.e., $(R, \Omega, a) = (550, 1.44, 40)$). For the relationship between the body-fitted and cylindrical coordinates, see section \ref{sec:2.2}.
(c) and (d) show the same plots as (a) and (b), respectively, but for $De_{\text{max}} = 50$.}
\label{fig:fig8}
\end{figure}

At the outlet $X=150$, the slow-scale effects have essentially vanished, and the solution approaches a travelling wave in a constant-curvature channel. However, since we are solving the reduced system (\ref{NSeq2}), it is of interest to assess the accuracy of this approximation.
To verify whether it provides a good approximation to the ECS of the full governing equations (\ref{NSeq}), we employ the numerical code developed by \cite{DEGUCHI_NAGATA_2011}, which computes ECS in concentric cylinders. The corresponding travelling wave solution of the full Navier-Stokes equations can be obtained in cylindrical coordinates by considering flow between two concentric cylinders with a mean radius of curvature $a=40$. 

Figure \ref{fig:fig8}-(a) shows the flow field of the outlet solution from the reduced system, while panel (b) presents that of the ECS solution in annulus.
Comparison is made at the same bulk Reynolds number and frequency.
Overall, the agreement is reasonable. The outlet solution has $\alpha=1.41$, which compare well with $1.44$ for the full Navier-Stokes solution (in cylindrical coordinates, $\alpha$ is defined as the azimuthal wavenumber multiplied by $a$). 
Note that some differences in the flow details are not surprising, as $De_{\text{max}}$ was chosen near the limit where $|\Pi'|$ remains close in value (less than 5\%) of both ECS. If $De_{\text{max}}$ is reduced, the agreement improves significantly, as shown in panels (c) and (d). At this level of curvature ($a=O(10^3)$), the errors in $(\alpha,-\Pi')$ are less than 0.1\%.


\subsection{Marching with MS-A2 as the inlet condition}\label{sec:4.2}

A qualitatively different result arises in the spatial marching problem when MS-A2 from table~\ref{table:inlet} is used as the inlet condition. 
The values of $\beta$ and $\Omega$ are set to 2.82 and 1.52, respectively, as listed in the table, while all other computational conditions are kept the same as in section~\ref{sec:4.1}.

\begin{figure}
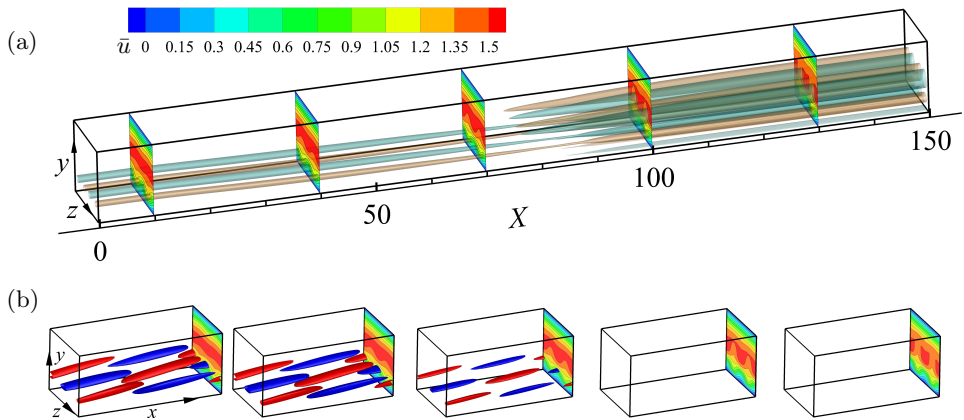

\centering
\begin{overpic}[width=0.9 \textwidth]{A/fig9/fig9.pdf}
\put(-4,40){(a)}
\put(-4,12){(b)}
\put(8,39.5){{$\bar u$}}
\end{overpic}
\caption{
Same computation and plot as figure \ref{fig:ND13_vis}, but with the inlet initial condition set to MS-A2
(the blue circle in figure \ref{fig:bd_R}, see table 1 also). Only a snapshot of the fluctuation field at one representative time is shown for simplicity.
We use the end-point value in figure~\ref{fig:fig6}-(a), $\alpha = 1.51$, to visualise the two rightmost boxes in panel (b).
} 
\label{fig:new_vis}
\end{figure}

\begin{figure}
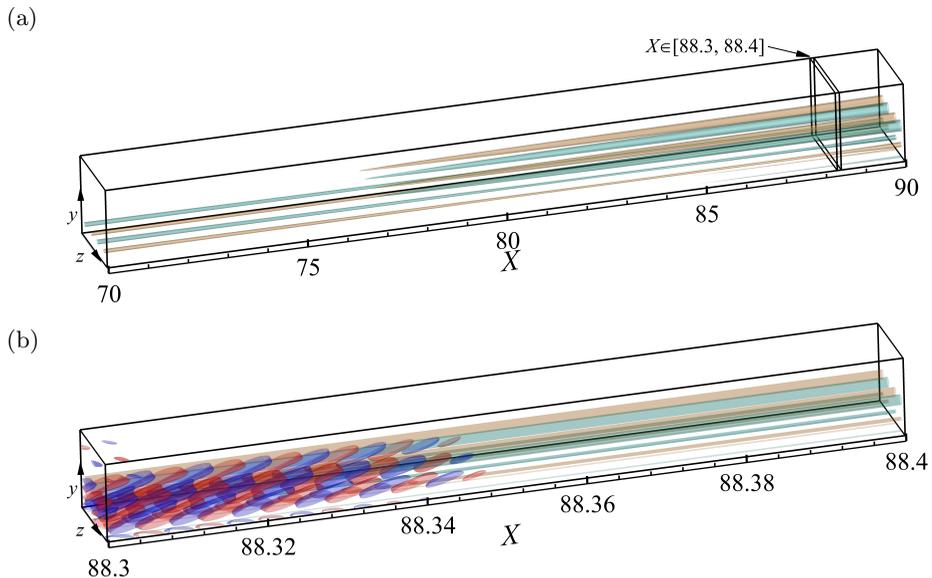

\centering
\begin{overpic}[width=0.9 \textwidth]{A/fig10/fig10.pdf}
\put(-4,60){(a)}
\put(-4,25){(b)}
\end{overpic}
\caption{
Detailed view of the flow field near the location where the fluctuation field vanishes in figure \ref{fig:new_vis}. (a) Only the mean field is shown. The brown/green surfaces depict 50\% maximum/minimum of the streamwise vorticity, $\partial_y\overline{w}-\partial_z \overline{v}$.
(b) Enlarged view of the small box indicated in panel (a), with the fluctuation field vorticity overlaid (red/blue isosurfaces are 10\% maximum/minimum of $\partial_y\tilde{w}-\partial_z \tilde{v}$).
%
} 
\label{fig:fig10}
\end{figure}
The result is shown in figure~\ref{fig:new_vis}. Panel (a) shows the mean field, which consists of the roll and streak structure enhanced by centrifugal instability; this is qualitatively similar to the previous case. However, as seen in panel (b), the fluctuation field disappears before reaching $X=100$. 
This transition occurs around $X = 88.35$, as highlighted in the close-up view (figure~\ref{fig:fig10}). Beyond this critical point, the computation reduces to the BRE formulation (\ref{eqBRE}), and  the value of $\alpha$ does not affect the flow field (see figure~\ref{fig:fig6}-(a)).

The solution for $X > 88.35$ can be interpreted as a Dean vortex solution, which does not depend on the fast scale. The mean field at the outlet $X = 150$ satisfies the approximate equation (\ref{NSeq2}) with $\partial_x=0$ and constant $De$, which is none other than the reduced system originally derived by Dean. We confirmed that the solution bifurcates from the second linear unstable mode of the laminar flow. The comparison with the full Navier-Stokes solution in the annulus, shown in figure \ref{fig:fig11}, confirms that the approximation is reasonably accurate, with the value of $|\Pi'|$ deviating by only 1 \%.


The blue dot-dashed curve in figure~\ref{fig:fig6}-(b) shows that emergence of Dean vortices increases the pressure gradient $|\Pi'|$. Nevertheless, it remains consistently smaller than the red curve obtained using MS-A. The inlet-outlet pressure difference driving the latter flow is 578.98, whereas that of the former is significantly lower at 496.06. 

\begin{figure}
\centering
\begin{overpic}[width=0.95 \textwidth]{A/fig11/fig11_new.pdf}
\put(-3,39.3){(a)}
\put(47,39.3){(b)}
\end{overpic}
\caption{Comparison between the reduced system solution and the Navier–Stokes solution in the constant-curvature case corresponding to the case shown in figure \ref{fig:new_vis}. The format of the figures is the same as figure \ref{fig:fig8}. 
(a) is the Dean vortex solution taken near the outlet $X=150$. (b) shows the steady axisymmetric Navier–Stokes solution in the annulus, maintaining the same bulk Reynolds number.
}
\label{fig:fig11}
\end{figure}

\begin{figure}
\centering
\begin{overpic}[width=0.9 \textwidth]{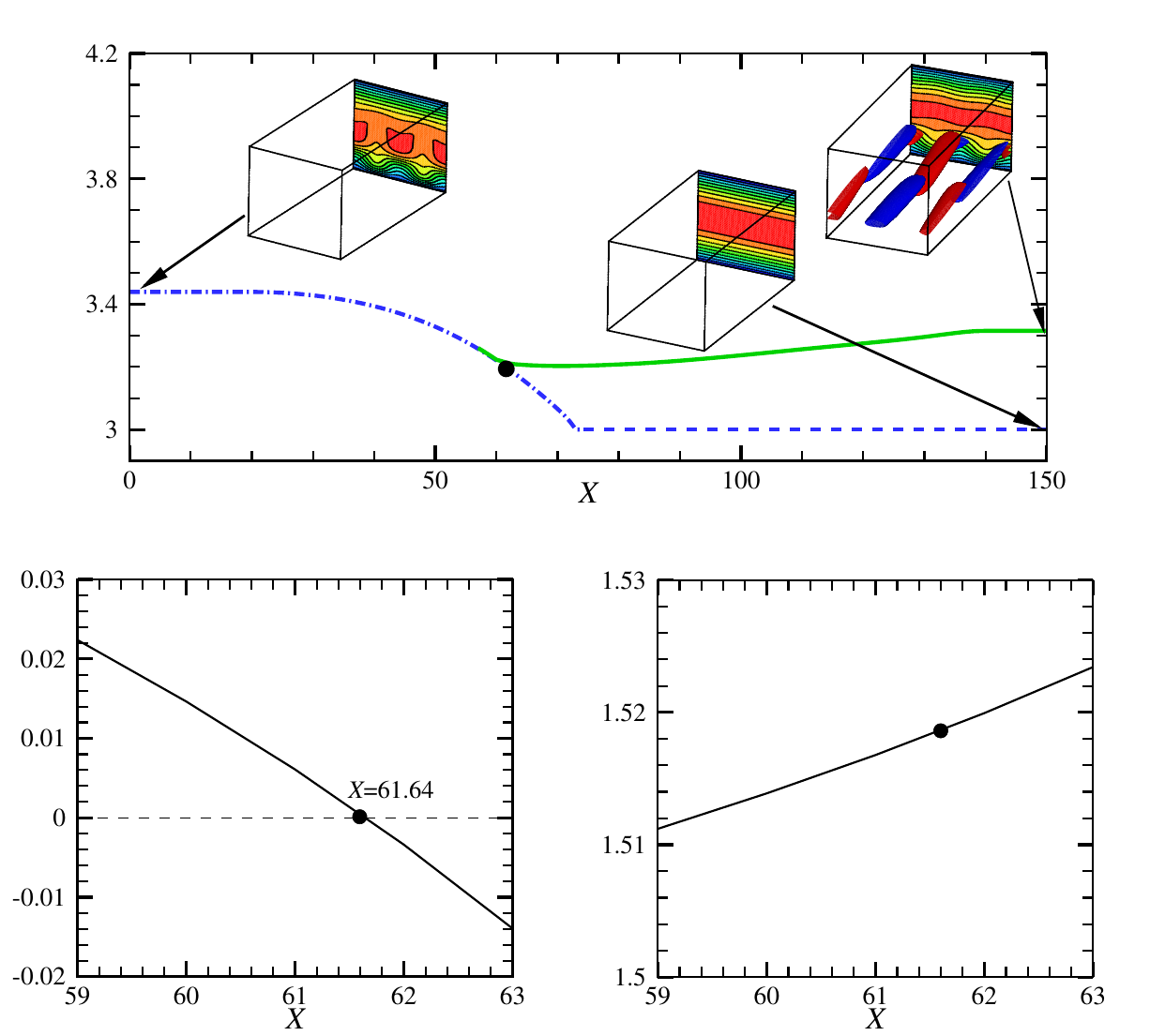}
\put(2,62){\begin{turn}{90}{$-\Pi'$}\end{turn}}
\put(2,84){(a)}
\put(-3,38){(b)}
\put(0,19){\begin{turn}{90}{$\alpha c_i$}\end{turn}}
\put(47,38){(c)}
\put(50,19){\begin{turn}{90}{$\Omega$}\end{turn}}
\end{overpic}
\caption{
The results of spatial marching computation with the inlet and outlet of the curved channel reversed. 
(a) The outcome of the spatial marching computation with the inlet condition chosen as the Dean vortex solution shown in figure \ref{fig:fig11}-(a). Along the dot-dashed line
the Dean vortices are sustained, while the dashed line represents the unidirectional laminar solution. The green solid line corresponds to the result with a small external forcing of frequency 1.518.
(b) Growth rate obtained from the secondary stability analysis of Dean flow. (c) The corresponding frequency of the eigenmode.
%
}
\label{fig:fig12}
\end{figure}
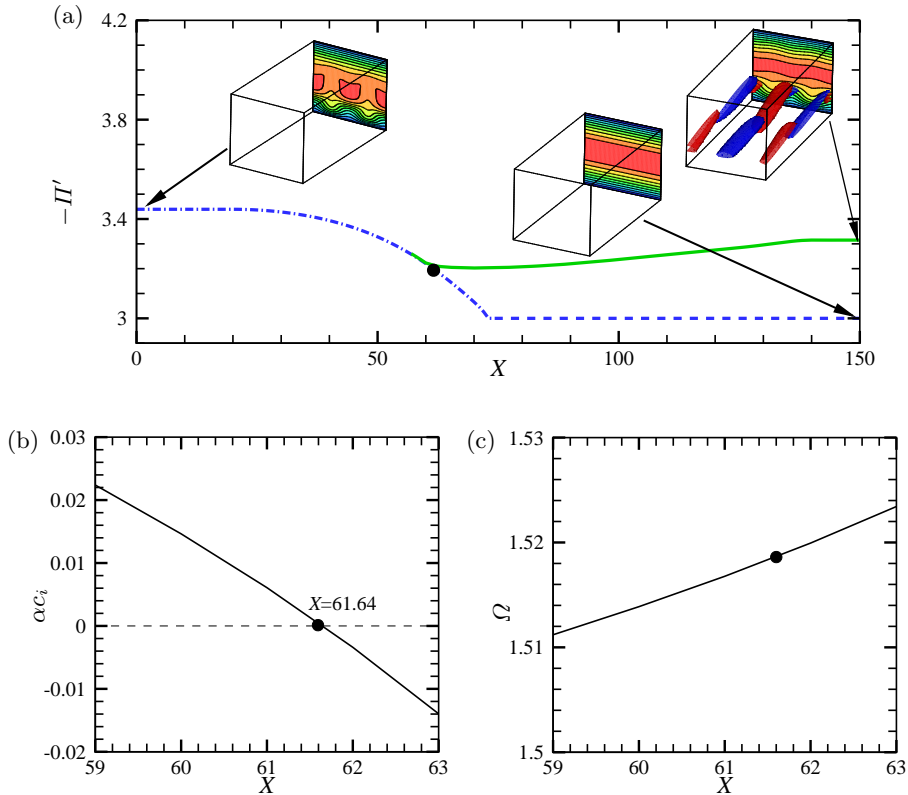

\subsection{Transition from Dean vortices to self-sustained states}\label{sec:4.3}

In section~\ref{sec:4.2}, the numerical computation was initiated from a self-sustained ECS of plane Poiseuille flow, MS-A2, and resulted in a Dean vortex solution at the outlet, $X = 150$.
A natural question then arises: what happens when the Dean vortex solution is used as the initial condition in a system with reversed inlet and outlet? Is it possible to recover MS-A2 after the channel becomes straight? 

The answer to this question is no, if the spatial marching is initiated from the Dean vortex solution shown in figure \ref{fig:fig11}.
The result of the spatial marching is shown by the blue line in figure~\ref{fig:fig12}-(a). 
Here and hereafter, we switch to the curvature $De(X) = De_{\text{max}}\,\chi(150 - X)$, while keeping the inlet position at $X=0$.
As the curvature decreases with $X$, the Dean vortices (dot-dashed line) gradually weaken and vanish beyond $X\approx 70$ 
(the unidirectional laminar flow, indicated by the dashed line, has $\Pi'=-3$).
%

However, with a subtle adjustment, a self-sustained state can emerge as the final state, as shown by the green line in figure~\ref{fig:fig12}-(a). Two steps are required to obtain this result. 
We first note that when the fluctuation field is infinitesimally small, the fluctuation equation (\ref{waveeqq}) can be linearised with respect to that field. Writing 
$\tilde{\mathbf{u}}=e^{i\theta}\hat{\mathbf{u}}(X,y,z)$ and $\tilde{p}=e^{i\theta}\hat{p}(X,y,z)$, the linearised equations become
\begin{subequations}\label{secondary}
\begin{eqnarray}
\mathscr{L}\hat{u}+\hat{v}\partial_y \overline{u}+\hat{w}\partial_z \overline{u}=-i\alpha \hat{p},\\
\mathscr{L}\hat{v}+\hat{v}\partial_y \overline{v}+\hat{w}\partial_z \overline{v}=-\partial_y \hat{p},\\
\mathscr{L}\hat{w}+\hat{v}\partial_y \overline{w}+\hat{w}\partial_z \overline{w}=-\partial_z \hat{p},\\
i\alpha \hat{u}+\partial_y\hat{v}+\partial_z\hat{w}=0,
\end{eqnarray}
\end{subequations}
where $\mathscr{L}=i\alpha(\overline{u}-c)+\overline{v}\partial_y+\overline{w}\partial_z-R^{-1}(\partial_y^2+\partial_z^2-\alpha^2)$.
The above set of equations represents the temporal secondary stability problem under the parallel-flow approximation, a well-established approach in boundary layer theory; see \cite{Hall_Horseman_1991,Li_Malik_1995,Xu_Zhang_Wu_2017}, for example (note that \cite{Hall_Horseman_1991} used (\ref{Rayleigh})).
At each $X$, the Dean vortex is substituted into $\overline{\mathbf{u}}$ in (\ref{secondary}), and the equations are solved for the eigenvalue $c=c_r+ic_i$. The result with $\alpha=1.51$ is shown in figures~\ref{fig:fig12}-(b) and (c); the value of $\alpha$ is taken from the point where the blue curve in figure~\ref{fig:fig6}-(a) terminates. 
The location where the growth rate becomes zero is at $X=61.64$, and the corresponding frequency at that point is 
1.518.

\begin{figure}
\centering
\begin{overpic}[width=0.95 \textwidth]{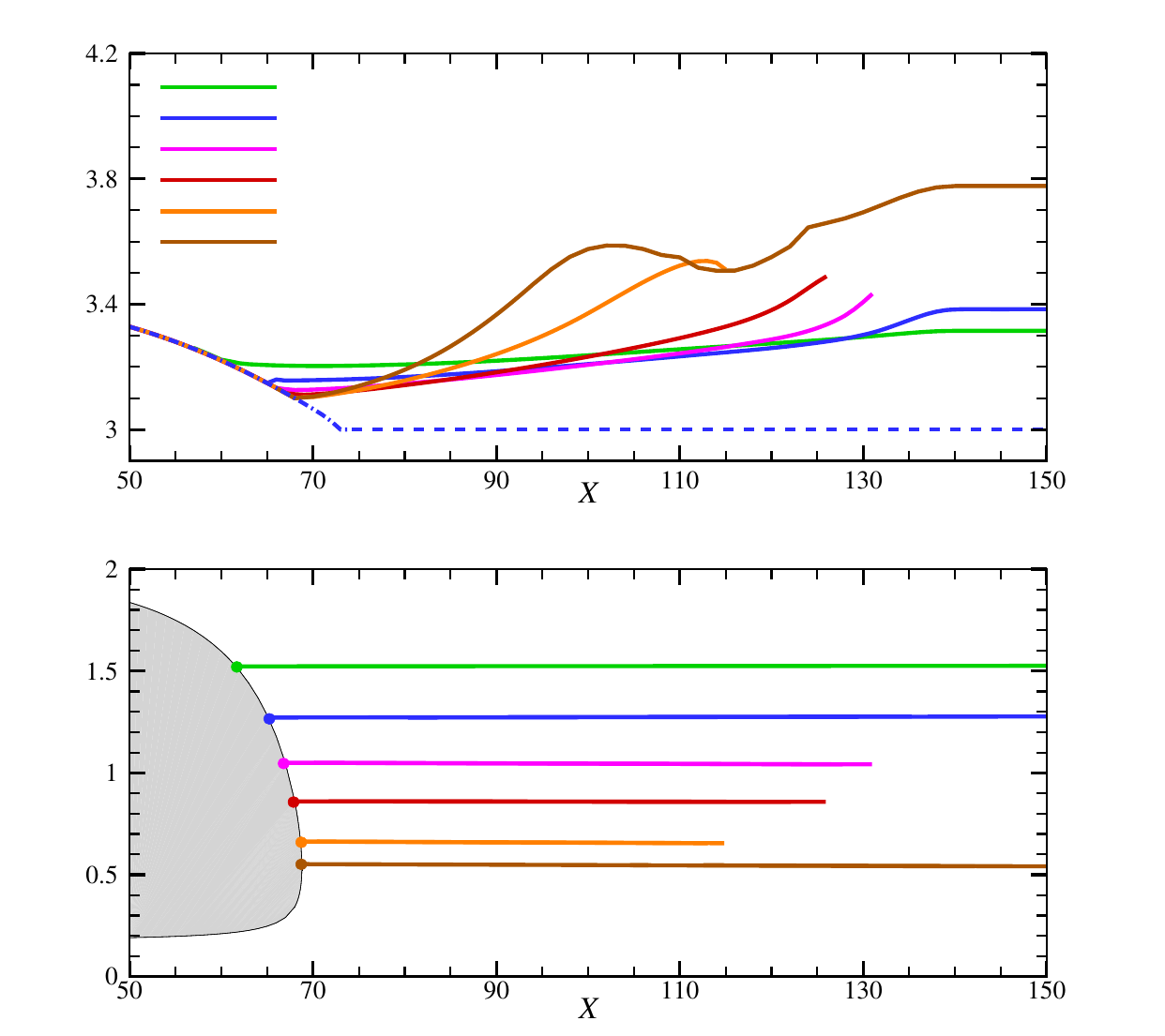}
\put(2,83){(a)}
\put(2,39){(b)}
\put(5,65){\begin{turn}{90}{$-\Pi'$}\end{turn}}
\put(5,22){\begin{turn}{90}{$\Omega$}\end{turn}}
\put(27,32.5){{$\Omega=1.518$}}
\put(27,28){{$\Omega=1.28$}}
\put(27,23.9){{$\Omega=1.05$}}
\put(27,20.4){{$\Omega=0.86$}}
\put(27,17){{$\Omega=0.66$}}
\put(27,12.5){{$\Omega=0.55$}}
\put(25,80.8){{$\Omega=1.518$}}
\put(25,78){{$\Omega=1.28$}}
\put(25,75.2){{$\Omega=1.05$}}
\put(25,72.4){{$\Omega=0.86$}}
\put(25,69.8){{$\Omega=0.66$}}
\put(25,67.4){{$\Omega=0.55$}}
\end{overpic}
\caption{
(a) Results of applying external forcing with various frequencies to the same marching problem as in figure \ref{fig:fig12}-(a). 
(b) The shaded region indicates the linearly unstable regime predicted by the stability analysis of the Dean vortex solution. The horizontal lines correspond to the solutions shown in panel (a).}
\label{fig:fig13}
\end{figure}
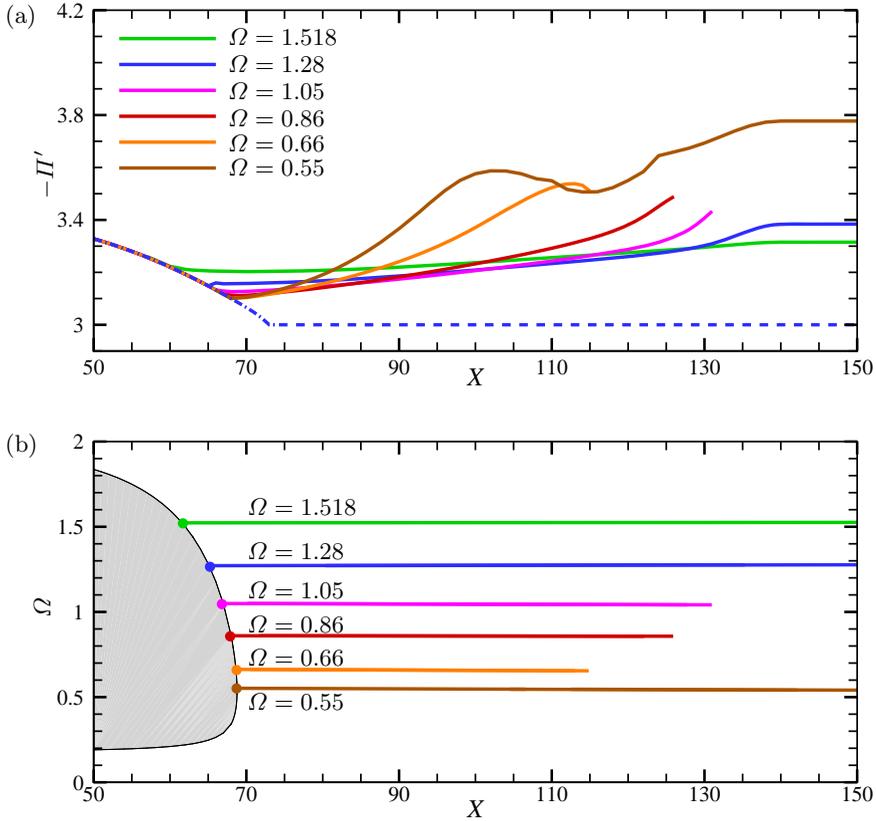

\begin{figure}
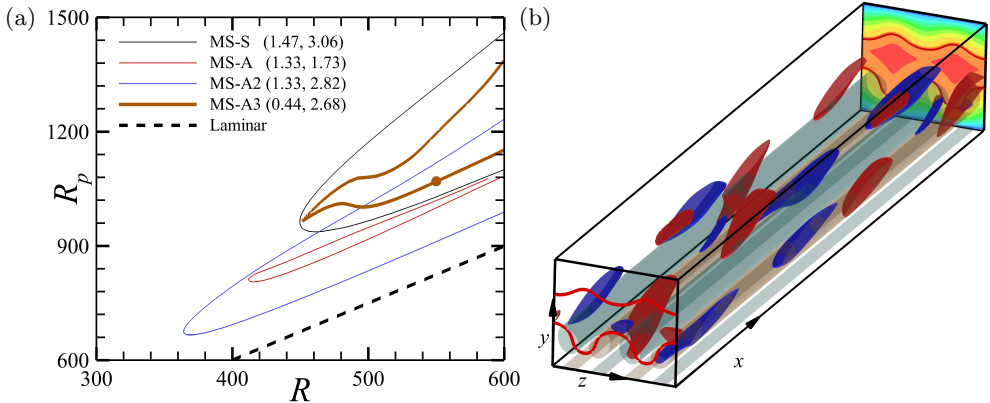

\centering
\begin{overpic}[width=0.95 \textwidth]{A/fig14/fig14.pdf}
\put(-4,39.3){(a)}
\put(49,39.3){(b)}
\end{overpic}
\caption{
The MS-A3 solution.
(a) Bifurcation diagram shown in the same format as figure \ref{fig:bd_R}. (b) Flow visualisation of the lower branch MS-A3 solution at $R=550$ (circle in panel (a)).
See the caption of figure \ref{fig:fig5} for  definitions of the isosurfaces and the colour map.
%
} 
\label{fig:fig14}
\end{figure}

We then add a small random external forcing term with $\theta$ dependence to (\ref{waveeq}). This forced marching problem is initiated from the Dean vortex at $X=60$, which is slightly upstream of the neutral point. The value of forcing frequency $\Omega=1.518$ is chosen to match the one obtained at the neutral point. The result corresponds to the green curve in figure~\ref{fig:fig12}-(a) mentioned earlier. The solution at $X=150$ converges to MS-A2, although the value of $\alpha$ differs slightly from that shown in table~\ref{table:inlet}.

If the local parallel approximation were used by setting $\partial_X=0$, the above computation could be interpreted as an imperfect bifurcation, as the forcing breaks the $\theta$-invariance of the Dean vortices. However, since we solve the evolution equations along the streamwise direction, this interpretation is not exact. Instead, the problem is more closely related to the receptivity problem in aeronautics \citep{Morkovin_1969,saric_2002}, which concerns the relationship between the input (external forcing from wall roughness or freestream perturbations) and the output (downstream perturbations) in a linear system. Near the neutral point, our analysis can be interpreted as a nonlinear extension of this framework. Another important distinction from conventional boundary layer transition is that, in our case, the base flow (Dean vortex) is initially unstable in the upstream region, and receptivity becomes most effective when the flow stabilises. To the best of our knowledge, such ‘subcritical’ receptivity has not been explored before.

Similar computations can be performed for various forcing frequencies, as shown in figure \ref{fig:fig13}-(a). 
Panel (b) shows the frequency range where instability occurs. When the frequency $\Omega$ falls below approximately 1.2, the marching terminates before reaching the outlet due to the failure of the Newton method to converge. 
Interestingly, however, once $\Omega$ is decreased further to around 0.55, the marching suddenly becomes stable and can proceed all the way to the outlet.

The plane Poiseuille solution found at this frequency is qualitatively different from any of the solutions introduced in section~\ref{sec:3.2}, and we therefore refer to it as MS-A3. 
This difference can be seen from the bifurcation diagram in figure \ref{fig:fig14}-(a) and the flow field in figure \ref{fig:fig14}-(b), both presented in a format that allows direct comparison with figures \ref{fig:bd_R} and \ref{fig:fig5}, respectively. These computations here thus demonstrate that the marching method can serve as a practical tool for discovering new ECS in parallel flows.











\section{Concluding remarks}\label{sec:5}

We have extended ECS-based analyses of coherent structures beyond simple parallel flows to configurations with slow spatial development. More specifically, we have developed an approach that enables efficient computation of time-periodic solutions in large domains, where full Navier–Stokes computations are prohibitive. Our main result is the derivation of the reduced system (\ref{vortexeq})–(\ref{waveeqq}), which couples the BRE with local ECS. Although approximate, the method retains all terms required by major nonlinear high-Reynolds-number asymptotic theories, as discussed in section \ref{sec:2.4}. 

The numerical implementation of the reduced problem also naturally combines the BRE formulation with the ECS-based approach (see section \ref{sec:3}). 
It is based on a standard multiple-scale analysis that incorporates non-parallel effects. The flow field is assumed to be periodic in the fast scale variable, which enables the use of a Newton-Raphson code originally developed for computing ECS in parallel flows, with modifications to include additional terms. Those additional terms arise from the implicit finite difference spatial marching scheme on the slow scale.
From a theoretical perspective, the fast-scale periodicity is essential to prevent the asymptotic expansion from breaking down over long times and distances. 
Note however that the constructed flow field is not locally periodic, due to the multiple-scale formulation, as explicitly demonstrated in section~\ref{sec:4}.

The proposed numerical method was applied to a curved channel flow with non-uniform curvature. In sections~\ref{sec:4.1} and \ref{sec:4.2} we examined a case in which the curvature gradually increases from zero to a value exceeding the linear critical Dean number predicted by the parallel flow approximation. A time-periodic finite-amplitude perturbation is imposed as the inlet condition, under the assumption that the ECS, corresponding to an edge state, is sustained upstream. Two types of ECS in plane Poiseuille flow, MS-A and MS-A2, were used as inlet conditions. Although both share the same symmetry, they produce qualitatively and quantitatively different flow fields downstream. When MS-A is used as the inlet condition, the mean field yields a steady, spatially developing vortex structure, whereas the fluctuation field persistently maintains wave-like coherent structures (section~\ref{sec:4.1}). On the other hand, in the case of MS-A2, the fluctuation field disappears partway through the domain, leaving behind Dean vortices that develop on the slow spatial scale (section~\ref{sec:4.2}). 
Interestingly, the two inlet conditions require significantly different pressure gradients (i.e. momentum transport) to sustain the corresponding global flow states.
This highlights the importance of controlling finite-amplitude inlet conditions to achieve power savings. In general, predicting the downstream outcome in the presence of finite-amplitude oncoming disturbances is inherently challenging, but our method may provide valuable guidance.

In section~\ref{sec:4.3}, we reversed the flow direction to investigate how the Dean vortices obtained in the MS-A2 case evolve as the channel curvature decreases. As expected, the Dean vortices decay and eventually vanish downstream. Nevertheless, by adding a small fast scale forcing, the self-sustaining process can be triggered, allowing a nontrivial solution to persist even after the curvature drops to zero. The effect of the forcing is most efficient when the Dean vortices become nearly neutral to secondary instability. This phenomenon is analogous to receptivity in boundary layer theory; in our case, an artificial external forcing was used, but it could be replaced with a physical forcing in future studies. 
From the perspective of reducing momentum transport, our analysis offers insight into which frequencies of external noises should be preferentially suppressed and at which locations.
It should be noted that our self-sustained state emerges subcritically, in the sense that it persists even as the secondary instability becomes stabilised.
Such a transition may also take place in boundary layer flows influenced by finite-amplitude vortical disturbances and weak external forcing, and could provide new insight into the nature of bypass transition. 







In practice, the inlet condition relevant to applied problems is most likely turbulent, and examining a single ECS is not sufficient. However, dynamical systems theory suggests that turbulence can be understood in terms of a collection of ECS. Therefore, using multiple inlet ECS in computations may eventually provide an efficient approximation of global turbulent flow fields.
We also note that subcritical transition is often characterised by the appearance of turbulent spots or bands. ECS resembling such localised structures have been reported in plane Poiseuille flow \citep{Brand_Gibson_2014,Paranjape_2023}, and it would be an interesting direction for future work to investigate how these structures are influenced by downstream channel curvature.



\backsection[Acknowledgements]{
This research was supported by the Australian Research Council Discovery Project DP230102188. }

\backsection[Declaration of Interests]{
The authors report no conflict of interest.
}

\appendix

\section{Navier-Stokes equations in the body-fitted coordinates}\label{app:A}

The Navier–Stokes equations (\ref{NSeq}), expressed in the body-fitted coordinates introduced in section~\ref{sec:2.1}, can be written as follows (see \cite{slattery1999}, for example):
\begin{subequations}\label{Bodyfitgoveq}
\begin{eqnarray}
\mathcal{D}u - \kappa \frac{uv}{H}=-\frac{p_x}{H}+\frac{1}{Re}\left \{\mathcal{L}u-\kappa (\frac{u}{H})_y-\frac{\kappa v_x}{H^2}-\frac{1}{H}(\frac{\kappa v}{H})_x \right\},\label{BodyU}\\
\mathcal{D}v+\frac{\kappa u^2}{H}=-p_y+\frac{1}{Re}\left \{\mathcal{L}v-\kappa (\frac{v}{H})_y+\frac{\kappa u_x}{H^2}+\frac{1}{H}(\frac{\kappa u}{H})_x \right \},\label{BodyV}\\
\mathcal{D}w=-p_z+\frac{1}{Re}\left \{\mathcal{L}w-\frac{\kappa}{H}w_y\right \},\\
\frac{1}{H}u_x+v_y-\frac{\kappa v}{H}+w_z=0.
\end{eqnarray}
\end{subequations}
Here, $H=1-\kappa y$ is the Lam\'e coefficient, and $\kappa (x)=1/a$ denotes the curvature of the channel centreline. To simplify the notation, we used subscripts for partial derivatives and defined the operators
\begin{subequations}
\begin{eqnarray}
\mathcal{D}f = f_t + \frac{u}{H} f_x + v f_y + w f_z,\qquad 
\mathcal{L}f = \frac{1}{H} \left( \frac{f_x}{H} \right)_x + f_{yy} + f_{zz}.
\end{eqnarray}
\end{subequations}
The analysis in section~\ref{sec:2.4} proceeds by substituting the mean–fluctuation decomposition (\ref{decomp}) and the asymptotic expansions (\ref{waveVWIexp})
and
(\ref{vortexVWIexp}) into the equations above.

From the definition of the Dean number, we have $\kappa=De^2/8R^2$, implying that when $R$ is large, $H=1+O(R^{-2})$. 
Thus, among the terms in (\ref{Bodyfitgoveq}), most of those absent in the Cartesian form of the Navier–Stokes equations are unimportant. However, the second term on the left-hand side of (\ref{BodyV}) requires careful attention. 
To see this, let us consider the selected terms
\begin{eqnarray}
\frac{u}{H}v_x+vv_y+\kappa \frac{u^2}{H}.\label{three}
\end{eqnarray}
By substituting the expansions
(\ref{vortexVWIexp})
and
(\ref{waveVWIexp}), we obtain the $\theta$-mean part
\begin{eqnarray}
R^{-2}(UV_X+VV_y+\frac{De^2}{8}U^2)\hspace{50mm}\nonumber\\
+\delta R^{-2}\{(\alpha\hat{u}\hat{v}^{\dagger}_{\theta}+\hat{v}\hat{v}^{\dagger}_y)+(R^{-1}\hat{u}\hat{v}_X^{\dagger}+R^{-2}\frac{De^2}{8}\hat{u}\hat{u}^{\dagger})\}+\text{c.c.}+\cdots\label{meanAp}
\end{eqnarray}
and the $\theta$-fluctuation part 
\begin{eqnarray}
\delta^{1/2}R^{-1} \{i\alpha U \hat{v}+R^{-1}(U\hat{v}_X+V\hat{v}_y+V_y\hat{v})\nonumber \\
 +R^{-2}(V_X\hat{u}+\frac{De^2}{4}U\hat{u}+i\alpha U \hat{v}\frac{De^2}{8}y)\}e^{i\theta}+\text{c.c.}+\cdots.~~\label{flucAp}
\end{eqnarray}
The dagger symbol denotes the complex conjugate.

The first group of terms in (\ref{meanAp}) corresponds to the BRE terms, while the second group represents the Reynolds stress. In a formal asymptotic expansion, only the former appears at leading order outside the critical layer.
To derive the reduced system (\ref{vortexeq}), we use assumption (iii) to discard the second group of terms within the curly brackets. (The first group must be retained, as it becomes leading order within the critical layer; see the discussion in section~\ref{sec:2.4}).

Inside the curly brackets in (\ref{flucAp}), there are terms proportional to $R^0, R^{-1}$, and $R^{-2}$. 
The $O(R^0)$ term appears in the Rayleigh equation (\ref{Rayleigh}) and is clearly the only term retained in the asymptotic analysis. 
In the derivation of the reduced system (\ref{waveeqq}), we also retain the second and third terms within the $O(R^{-1})$ group, while the first term is discarded in accordance with assumption (iv) (see section~\ref{sec:2.3}). By the same assumption, the first and second terms in the $O(R^{-2})$ group can be neglected.
The last term in the $O(R^{-2})$ group originates from the $1/H$ factor in the first term of equation (\ref{three}). This term does not appear in the reduced system due to assumption (i).

\bibliographystyle{jfm}  
\bibliography{Reference}  
\end{document}